\documentclass[10 pt, letterpaper]{article}
\usepackage[T1]{fontenc}
\usepackage{times}
\usepackage{amsmath}
\usepackage{amssymb}
\usepackage{graphicx}
           \onecolumn

\newtheorem{assumption}{Assumption}

\newtheorem{algorithm}{Algorithm}

\newtheorem{remark}{Remark}

\def\bt{\boldsymbol{\theta}}
\def\ba{\boldsymbol{\alpha}}
\def\baa{\boldsymbol{a}}
\def\bx{\boldsymbol{x}}
\def\bv{\boldsymbol{v}}
\def\bw{\boldsymbol{w}}
\def\bc{\boldsymbol{c}}

\def\bP{\boldsymbol{\Phi}}
\def\P{\Phi}
\def\t{\theta}
\def\a{\alpha}
\def\s{\sigma}
\def\R{\mathbb{R}}
\def\N{\mathbb{N}}

\begin{document}
\date{}
\title{Simulation of stochastic systems via polynomial chaos expansions and convex optimization\thanks{This manuscript is a preprint of a paper published on Physical Reviews E and is subject to American Physical Society copyright. The copy of record is available at http://pre.aps.org. URL: http://link.aps.org/doi/10.1103/PhysRevE.86.036702; DOI: 10.1103/PhysRevE.86.036702}
\thanks{This research has received funding from the European Union Seventh Framework
Programme (FP7/2007-2013) under grant agreement n. PIOF-GA-2009-252284 -
Marie Curie project ``ICIEMSET'', from the Air Force Office of Scientific Research under the MURI award n. FA9550-10-1-0143, from the  National Science Foundation through Grants  ECCS-0835847 and ECCS-0802008 and from the Institute for Collaborative Biotechnologies through Grant DAAD19-03-D-0004 from the US Army Research Office.}}

\author{Lorenzo Fagiano\thanks{Dip. di Automatica e Informatica, Politecnico di Torino,
Italy, and  Dept. of Mechanical Engineering, University of California at Santa Barbara, USA. E-mail: lorenzo.fagiano@polito.it.}$\,$ and Mustafa Khammash\thanks{Dept. of Mechanical Engineering, University of California at Santa Barbara, USA, and Dept. of Biosystems Science and Engineering, ETH Zurich, Switzerland. E-mail: mustafa.khammash@bsse.ethz.ch.}
}
\maketitle

\noindent\textbf{{Abstract} --
Polynomial Chaos Expansions represent a powerful tool to simulate stochastic models of dynamical systems. Yet, deriving the expansion's coefficients for complex systems might require a significant and non-trivial manipulation of the model, or the computation of large numbers of simulation runs, rendering the approach too time consuming and impracticable for applications with more than a handful of random variables. We introduce
 a novel computationally tractable technique for computing the coefficients of polynomial chaos expansions.  The approach exploits a regularization technique with a particular choice of weighting matrices, which allow to take into account the specific features of Polynomial Chaos expansions. The method, completely based on convex optimization, can be applied to problems with a large number of random variables and uses a modest number of Monte Carlo simulations, while avoiding model manipulations. Additional information on the stochastic process, when available, can be also incorporated in the approach by means of convex constraints. We show the effectiveness of the proposed technique in three applications in diverse fields, including the analysis of a nonlinear electric circuit, a chaotic mo\-del of organizational behavior, finally a chemical oscillator.}\\
$\,$\\

\section{Introduction}\label{S:Intro}
In most science and engineering applications, there is the need to simulate mathematical models of the process under study, in the form of ordinary or partial differential equations, with the aim to figure out the time (or space) course of a variable of interest $v$ for analysis, decision-making and control. In many cases, in order to take into account the presence of uncertainty, unknown external inputs and in general any effect that produces a mismatch between the model equations and reality, the model to be simulated is not fully deterministic: uncertainty and disturbance are often modeled as quantities with stochastic nature, named here ``input random variables'' and indicated with $\bt$, hence the name stochastic models. In these cases, $v(\bt)$ is a random variable, too, and one is interested in computing its statistics. The issue of simulating complex, nonlinear stochastic models with sufficiently high accuracy and low computational effort is still a challenge in important and diverse fields, like analysis of large power grids, weather forecasts at different scales and simulation of biological systems, to name just a few. The typical approach followed to simulate a stochastic model is the well-known Monte Carlo (MC) technique, which relies on the sampling of a finite number $M$ of values of $\bt$, according to its distribution. With sufficiently large $M$, the MC approach gives good statistical estimates (e.g. first and second order moments) of the variables of interest, and also of its probability density function (pdf). However, the application of MC simulations with the system model may be too computationally demanding, particularly in those cases when the model is complex and the inherent variables have large dimensions. Polynomial Chaos Expansions (PCEs) (see e.g. \cite{Wien38,CaMa47,GaSp91,XiKa02,WaKa05,XiHe05,WaRK09,GSVK10,BlSu11}) provide a useful tool to significantly reduce the computational effort required to simulate a stochastic system, by conceptually replacing the mapping between $\bt$ and $v$, implicitly defined by the integration of the model's differential equations, with an explicit function $\hat{v}(\bt)$, which takes the form of a truncated series of polynomials. The polynomials in the PCE are orthogonal, so that the statistical moments of $\hat{v}(\bt)$ can be computed directly from the expansion's coefficients. Moreover, the computational effort required to evaluate a PCE is often orders of magnitude lower than the one required to simulate the system model: therefore, it is possible to estimate the pdf of $v$ by using a Monte Carlo approach with the PCE, instead of the system model, with significant time savings. As an example, in the first case study considered in this paper, 100,000 Monte Carlo simulations with the system model require 3,800$\,$s, while the same number of PCE evaluations are obtained in 24$\,$s, with the same hardware and software. PCEs have been used with good results in a number of different areas, including experimental modeling, materials, mechanics, power systems, systems biology, and control, see e.g. \cite{Kaku50,BaCh96,GhGh02,XLSK02,LuKa03,HoLe04,LiSK04,HoTr06,KiDN07,Hove08,DaGF09,NaBr10,PeGL10}. Yet, the computation of the PCE's coefficient may not be a trivial task when the model is nonlinear and/or the dimension $n$ of random input variables is relatively high. In these cases, the existing approaches may require a significant and non-trivial manipulation of the model, or the computation of large numbers of simulation runs, such that the advantages of the method with respect to standard MC simulations might be lost. We propose here a novel approach to derive the PCE's coefficients, based on convex optimization. Under mild assumptions, this method can be easily applied to any existing model, since it just requires the preliminary computation of a small number of sampled values of $v$. In this approach, a relatively large order of the polynomial chaos is initially chosen, hence a high number of terms in the expansion, and the PCE's coefficients are then computed by means of a single multi-objective optimization problem, exploiting $\ell_1$ regularization techniques (see e.g. \cite{Tibshirani96},\cite{Donoho06}). We provide a new, systematic way, particularly suited to the properties of Polynomial Chaos Expansions, to choose the weighting matrices in the cost function of the optimization problem. Moreover, we show how different kinds of available information on the stochastic model, including bounds on $v$ and on its variance, can be easily taken into account as convex constraints in the optimization. As a result, the method is able to provide an accurate description of the statistics of $v(\bt)$, with few simulation runs. We present three case studies in a broad range of different fields, to demonstrate the effectiveness of the method and its ease of implementation.

\section{Problem Settings}\label{S:ProFor}
We consider a time-invariant system in state-space form:
\begin{equation}
\begin{array}{l}
\dfrac{d\bx(t)}{dt}=f\left(\bx(t),\bw(t),\bc,t\right) \\
\\
v=v(T)=h\left(\bx(T),\bw(T),\bc,t\right)
\end{array}
\label{sys1}
\end{equation}
where $t\in \R$ is the time variable, $\bx(t)\in \R^{n_x}$ is the system state,
$\bw(t)\in\R^{n_{w}}$ is an unknown input, $\bc\in\R^{n_c}$ is an unknown parameter vector, finally $v\in \R$ is a variable of interest, evaluated at a given time instant $T$. $\bw(t)$ and $\bc$ are assumed to have stochastic nature, in a sense that will be better detailed afterwards. Bold symbols indicate vectors of variables, e.g. $\bx=\left[x_1,\ldots,x_{n_x}\right]^T$, where $^T$ is the vector transpose operation. The aim is to derive an approximation of the first and second order moments and the pdf of  $v$, starting from (possibly stochastic) initial conditions $\bx(0,\bc)$, by using the model \eqref{sys1}. The variable $\bc$ accounts for uncertainty both in the model equations (e.g. due to uncertain physical parameters) and in the initial state $\bx(0,\bc)$, while $\bw(t)$ accounts for unknown external inputs, like disturbances.
The parameters $\bc$ and the unknown input $\bw(t),\,t\in[0,\,T]$ in \eqref{sys1} can be typically expressed as functions of a $n$-dimensional vector $\bt\in K^{n}$ of independent and identically distributed (iid) random variables $\t_i$, with known pdf $f_\t$, such that $\t_i\in K\subset\mathcal{L}^2(\Omega,\mathcal{F},P),\,\forall i\in\{1,\ldots,n\}$. Here,  $(\Omega,\mathcal{F},P)$ is a probability space, $\Omega$ is the set of elementary events, $\mathcal{F}$ is the $\sigma-$algebra of the events and $P$ is the probability measure. The expectation (or first-order moment) of a generic random variable $\t:\mathcal{A}\rightarrow\R$ is denoted as $E\left[\t\right]\doteq\int_\Omega\t(\omega)dP(\omega)=\int_\mathcal{A}\t dF_\t$, where $F_\t(k)\doteq P\{\t<k\}$ is the probability distribution function of $\t$ over $\mathcal{A}$. $\mathcal{L}^2(\Omega,\mathcal{F},P)$ is the Hilbert space of all random variables $\t$ whose $L_2$-norm, $\|\t\|_2\doteq E\left[|\t|^2\right]^{1/2}$, is finite, where $|\cdot|$ denotes the absolute value. $K$ is a subspace of $\mathcal{L}^2(\Omega,\mathcal{F},P)$ that  contains only centered random variables (i.e. $\forall\t\in K,\,E\left[\t\right]=0$). Finally, the pdf of $\t$ is given by $f_\t(k)=dF_\t/dk$, and the variance (or second-order moment) of $\t$ is indicated as $\text{Var}(\t)\doteq E\left[(\t-E[\t])^2\right]=\sigma_\t^2$, where $\sigma_\t$ is the standard deviation of $\t$.\\
In many cases of practical relevance, the time-invariant parameters $\bc$ are naturally iid variables (e.g. when $\bc$ stands for some variation of a physical parameter of the system, that is uncertain due to production variability). If the variables $c_i,\,i\in\{1,\,n_c\}$ have different probability distributions, it is possible to map a standard (i.e. with zero mean and unit variance) distributed Gaussian random variable, indicated as $\mathcal{N}(0,1)$, to a random variable
with distribution function $F_c$ by the transformation $F_c^{-1}(\text{erf}(\mathcal{N}(0,1)))$, where erf is the Gaussian distribution function, see e.g. \cite{Papo91}. As regards the input $\bw(t),\,t\in[0,\,T]$, this can typically be modeled as a stochastic process or random field $\bw(t,\bt):\R\times K^{n}\rightarrow\R^{n_{w}}$, which can be represented as a finite series of $n$ iid random variables multiplied by deterministic functions $\hat{\bw}_i(t),\,i\in\{1,\,n\}$, i.e. $\bw(t,\bt)\simeq\hat{\bw}_0(t)+\sum\limits_{i=1}^{n}\hat{\bw}_i(t)\t_i$ (see for example \cite{BaCh02,DeBO01,BaLT03,LuKa03}).\\
We assume that the solution of the dynamical equations \eqref{sys1} in the time interval $[0,\,T]$ exists and it is unique almost surely, i.e. with probability one.
In the described context, the variable of interest is a random variable, $v(\bt)$, and we name the system \eqref{sys1} ``stochastic system''. We assume that $v(\bt)$ has finite variance:
\begin{assumption}\label{A:variance}(Finiteness of variance of $v(\bt)$)\\
$v(\bt)\in\mathcal{L}^2(\Omega,\mathcal{F},P)$.
\end{assumption}
Assumption \ref{A:variance} is typically satisfied in practical applications.\\
The problem of simulating a stochastic system may be very complex and the main technique employed so far in engineering applications is the well-known MC approach, which consists in the following steps:
\begin{algorithm}\label{A:MC}(Monte Carlo simulations)
\begin{enumerate}
  \item extract $M$  iid samples $\tilde{\bt}_{(r)},\,r\in\{1,\,M\}$, of $\bt$, according to its distribution;
  \item for each sample, compute (or numerically simulate) the solution of \eqref{sys1} and the corresponding value of $v(\tilde{\bt}_{(r)}),\,\forall r\in\{1,\,M\}$;
  \item analyze the statistics of the collected data.
\end{enumerate}
\end{algorithm}
\begin{remark}\label{R:multiple_v}
For simplicity and without loss of generality, we consider a scalar variable $v$ and a single value of $T$: multiple variables of interests $v_j,\,j\in\{1,\,n_v\}$ and time values $T_i,\,i\in\{1,\,N\}$ can be easily treated by considering each variable and each time instant separately from the others, as it is done in the case studies reported in this paper. In these cases, a single simulation of the system provides all of the corresponding samples $v_j(T_i,\tilde{\bt}_{(r)}),\,\forall j\in\{1,\,n_v\},\,\forall i\in\{1,\,N\}$.
\end{remark}
Although MC simulations proved to be very effective in many applications, the required computational times may be prohibitive in various cases, e.g. when a decision has to be taken in relatively little time on the basis of the simulations' outcome, or when repeated MC simulations have to be carried out to tune some input or parameter, or  when the simulation has to be embedded in a numerical optimization procedure (e.g. for optimal design or control of stochastic systems). Polynomial Chaos Expansion techniques (see e.g. \cite{GaSp91}) are able to significantly reduce the computational effort required by standard MC approaches, by replacing the simulation of a (possibly very complex) dynamical system with the evaluation of a static function $\hat{v}(\bt)\approx v(\bt)$. The main features of PCEs are recalled in the next Section.

\section{Polynomial Chaos Expansions}\label{S:GPC}
PCEs were first introduced by Wiener \cite{Wien38}, who considered Gaussian random variables $\bt$. Later on, Cameron and Martin \cite{CaMa47} showed one of the key properties of PCEs, namely their ability to uniformly approximate any random process with finite second-order moments. The polynomial chaos is an orthogonal basis of $\mathcal{L}^2(\Omega,\mathcal{F},P)$, hence any random variable $v(\bt)\in\mathcal{L}^2(\Omega,\mathcal{F},P)$ has the $L_2$-convergent expansion \cite{CaMa47}:
\begin{equation}\label{E:PCE_v}
v(\bt)=\sum\limits_{k=0}^\infty a_{k}\P_{\ba_{k}}(\bt),
\end{equation}
where the coefficients $a_k$ are given by $a_k=\frac{E\left[v(\bt)\P_{\ba_{k}}(\bt)\right]}{E\left[\P_{\ba_{k}}(\bt)^2\right]}$, and $\P_{\ba_{k}}=\P_{\ba_{k}}(\t_1,\ldots,\t_n)$ is the $k$-th multivariate polynomial in the series, corresponding to the $k$-th vector of indices, or ``multi-index'', $\ba_k=\left[\a_{1,k},\ldots,\a_{n,k}\right],\,\a_{i,k}\in\N$. More specifically, for a given vector of indices $\ba_k$, we have $\P_{\ba_k}(\bt)\doteq\prod\limits_{i=1}^{n}\P_{(\a_{i,k})}(\t_i)$, where $\P_{(\a_{i,k})}(\t_i)$ is the  univariate polynomial of degree $\a_{i,k}$, chosen according to the Askey scheme \cite{XiKa02}. As an example, Hermite polynomials are used with Gaussian input random variables. Table \ref{T:ortho_polynomials} shows the suitable orthogonal polynomials for different kinds of input random variables. The choice of the univariate polynomials is made in order to satisfy the orthogonality property:
\begin{equation}\label{E:orthogonality}
E\left[\P_{(\a_{j})}\,\P_{(\a_{i})}\right]=E\left[\P_{(\a_{i})}^2\right]\delta_{ij},
\end{equation}
where $\delta_{ij}=1$ if $i=j$ and 0 in any other case.
\begin{table}
  \centering
  \caption{Examples of orthogonal polynomials for different kinds of probability measure}\label{T:ortho_polynomials}
\begin{tabular}{ll}\hline
\textbf{Random variable} & \textbf{Polynomial basis}\\\hline
Gaussian& Hermite\\
Uniform& Legendre\\
Gamma& Laguerre\\
Beta& Jacobi\\
\end{tabular}
\end{table}
The coefficients of the univariate polynomials can be usually computed via a recursive equation, starting from the terms of degree 0 and 1. As an example, Legendre polynomials, which are orthogonal w.r.t. to the uniform probability distribution, can be obtained as:
\begin{equation}\label{E:Legendre_recursion}
\begin{array}{l}
\P_{(0)}(\t)=1\\
\P_{(1)}(\t)=\t\\
\P_{(\alpha+1)}(\t)=\dfrac{2\alpha+1}{\alpha+1}\t\,\P_{(\alpha)}(\t)-\dfrac{\alpha}{\alpha+1}\P_{(\alpha-1)}(\t).
\end{array}
\end{equation}
We denote with $l_k\doteq\sum\limits_{i=1}^{n}\a_{i,k}$ the sum of the indices in the multi-index $\ba_k$, and we assume that the ordering of the multivariate polynomials $\P_{\ba_{k}}$ in \eqref{E:PCE_v}, and of the related coefficients $a_{k}$, is such that $l_k\leq l_{k+1}$. For practical reasons, the series \eqref{E:PCE_v} is truncated by considering only the multi-indices up to a maximal total degree $\overline{l}$, i.e. $\forall\ba_k:l_k\leq\overline{l}$. An example of ordering of all the multivariate polynomials corresponding to $\overline{l}=2,\,n=3$ is shown in Table \ref{T:expansion_example}. It can be clearly noted that the number of terms in the truncated series grows rapidly with $n$ and $\overline{l}$.
\begin{table}
  \centering
  \caption{Example of multivariate polynomials used in polynomial chaos, corresponding to $\overline{l}=2,\,n=3$}\label{T:expansion_example}
\begin{tabular}{cll}\hline
\textbf{Order} & \textbf{Multi-index}& \textbf{Multivariate Polynomial}\\\hline
0&$\ba_0=\left[0,0,0\right]$&$\P_{\ba_{n,0}}(\bt)=1$\\
1&$\ba_1=\left[1,0,0\right]$&$\P_{\ba_{n,1}}(\bt)=\P_{(1)}(\t_1)$\\
1&$\ba_2=\left[0,1,0\right]$&$\P_{\ba_{n,2}}(\bt)=\P_{(1)}(\t_2)$\\
1&$\ba_3=\left[0,0,1\right]$&$\P_{\ba_{n,3}}(\bt)=\P_{(1)}(\t_3)$\\
2&$\ba_4=\left[2,0,0\right]$&$\P_{\ba_{n,4}}(\bt)=\P_{(2)}(\t_1)$\\
2&$\ba_5=\left[0,2,0\right]$&$\P_{\ba_{n,5}}(\bt)=\P_{(2)}(\t_2)$\\
2&$\ba_6=\left[0,0,2\right]$&$\P_{\ba_{n,6}}(\bt)=\P_{(2)}(\t_3)$\\
2&$\ba_7=\left[1,1,0\right]$&$\P_{\ba_{n,7}}(\bt)=\P_{(1)}(\t_1)\P_{(1)}(\t_2)$\\
2&$\ba_8=\left[1,0,1\right]$&$\P_{\ba_{n,8}}(\bt)=\P_{(1)}(\t_1)\P_{(1)}(\t_3)$\\
2&$\ba_9=\left[0,1,1\right]$&$\P_{\ba_{n,9}}(\bt)=\P_{(1)}(\t_2)\P_{(1)}(\t_3)$\\
\end{tabular}
\end{table}
Since all the possible multi-indices $\ba$ that sum up to $l\leq\overline{l}$ are considered, the total number  $L$ of terms in the truncated expansion is:
\begin{equation}\label{E:N_calc}
L=\dfrac{\left(n+\overline{l}\right)!}{n!\,\overline{l}!},
\end{equation}
and the series takes the form:
\begin{equation}\label{E:PCE_v_trunc}
\hat{v}(\bt)\doteq\sum\limits_{k=0}^{L-1} a_{k}\P_{\ba_{k}}(\bt)=\bP(\bt)\baa,
\end{equation}
where $\baa\doteq[a_{0},\ldots,a_{L-1}]^T$ and
\begin{equation}\label{E:pol_vec}
\boldsymbol{\P}(\bt)\doteq[\P_{\ba_{0}}(\bt),
\ldots,\P_{\ba_{L-1}}(\bt)]
 \end{equation}
are, respectively, the vectors of the PCE's coefficients and of the multivariate polynomials evaluated at $\bt$. We refer to the truncated expansion $\hat{v}(\bt)\approx v(\bt)$ \eqref{E:PCE_v_trunc} as the PCE of the random variable $v(\bt)$.
The PCE has been shown to converge exponentially in the $L_2$-sense as the maximal order $\overline{l}$ increases, see e.g.  \cite{GaSp91,XiKa02}.
By applying the orthogonality property \eqref{E:orthogonality}, the first and second order moments of the random variable $\hat{v}(\bt)$ can be computed directly from the coefficients of its PCE, as follows:
\begin{equation}\label{E:PCE_moment_1}
\begin{array}{l}
E\left[\hat{v}(\bt)\right]=a_0
\end{array}
\end{equation}
\begin{equation}\label{E:PCE_moment_2}
\begin{array}{l}
\hat{\s}^2(\baa)\doteq\text{Var}\left(\hat{v}(\bt)\right)=\sum\limits_{k=1}^{L-1} a_k^2E\left[\P_{\ba_{k}}(\bt)^2\right]
\end{array}
\end{equation}
where $a_0$ is the coefficient of the polynomial of order $l=0$ (i.e. $\P_{\ba_{0}}=1$) in the PCE.
As regards the practical computation of equation \eqref{E:PCE_moment_2}, the terms $E\left[\P_{\a_k}(\bt)^2\right],\,\forall k\in\{1,L-1\}$ have to be computed once for all uses, and they can typically be obtained quite easily. As an example, for Legendre polynomials and uniformly distributed input random variables, note that, by considering that $f_\t=0.5d\t$, the $L_2$-norm squared of the multivariate polynomials is:
\begin{equation}\label{E:Legendre_norm}
E\left[\P_{\ba_k}(\bt)^2\right]=\|\P_{\ba_k}(\bt)\|_2^2=\prod\limits_{i=1}^{n}\dfrac{1}{2\alpha_{i,k}+1},\,\forall k\in\{0,L-1\}.
\end{equation}
Similar equations can be derived for the other types of orthogonal polynomials.\\
Moreover, a Monte Carlo approach can be used to estimate the pdf of $\hat{v}(\bt)$ (and, hence, of $v(\bt)$) once the coefficients of its PCE  are known, by simply evaluating the PCE $\hat{v}(\bt)$, instead of simulating the model \eqref{sys1} at step 2) of Algorithm \ref{A:MC}. The computational time required to evaluate the PCE is often orders of magnitude smaller than the one required to integrate numerically  the model \eqref{sys1}, hence the advantage of using polynomial chaos.\\
Clearly, one of the crucial points in the use of PCEs for the simulation of stochastic systems is the computation of the expansion's coefficients, $\baa$. In the literature, this task is carried out essentially in two different ways. A first method (see e.g. \cite{XiKa02}) relies on  a Galerkin projection to obtain an augmented set of deterministic differential equations, which can be solved to compute the PCE coefficients. While this method is quite attractive from a theoretical point of view, it might be affected by some practical issues, since for complex nonlinear models it may be difficult and too time-consuming to derive the augmented set of differential equations, and the number of such equations may be too large to obtain an efficient numerical solution with standard ODE solvers.\\
A second approach is known as Probabilistic Collocation Method (PCM, see e.g. \cite{XiHe05,NaBr10}), and it basically consists in the estimation of the coefficients from a finite number of data, i.e. of $\nu$ exact values of $v$, named ``collocation points'', corresponding to  $\nu$ values of the input random variables, $\tilde{\bt}_{(r)},\,r\in\{1,\,\nu\}$. Here, we consider a PCM-like approach for the computation of PCEs, since it appears to be more viable for the analysis of large-scale, complex stochastic dynamical systems, and we propose a new method to estimate the coefficients. The method and its features are described in the next section. One of the main advantages of probabilistic collocation is that no modification to the original model \eqref{sys1} is required, but just a series of preliminary simulations to collect the data to be used in the coefficients' computation; one of its main disadvantages is that the number of collocation points can be very high, for problems with relatively high stochastic dimensions (i.e. high values of $n$) and strong nonlinearities. Here, we will show, through a series of case studies, that our method yields very good results even with a very low number of collocation points.

\section{Computation of Polynomial Chaos Expansions via convex optimization}\label{S:SMOPCE}

Given the maximal order $\overline{l}$ of the PCE and the corresponding number of terms $L$ \eqref{E:N_calc}, we propose the following algorithm to estimate the PCE's coefficients $\baa$:
\begin{algorithm}\label{A:SPCE} (PCE computation via convex optimization)
\begin{enumerate}
  \item sample a finite number $\nu$ of independent values of the vector of input random variables $\tilde{\bt}_{(r)},\,r\in\{1,\,\nu\}$, according to its distribution;
  \item carry out $\nu$ simulations of the system \eqref{sys1}, each one corresponding to one of the extracted samples $\tilde{\bt}_{(r)}$;
  \item collect the obtained values of the variables of interest in the vector $\tilde{\bv}\doteq[v(\tilde{\bt}_{(1)}),\ldots,v(\tilde{\bt}_{(\nu)})]^T$;
  \item select the maximal order $\overline{l}$ for the PCE $\hat{v}(\bt)$ \eqref{E:PCE_v_trunc} and compute the matrix \[
      \tilde{\bP}\doteq\left[
      \begin{array}{c}
      \boldsymbol{\P}(\tilde{\bt}_{(1)})\\
      \vdots\\
      \boldsymbol{\P}(\tilde{\bt}_{(\nu)})
      \end{array}\right],\] where the vectors $\boldsymbol{\P}(\tilde{\bt}_{(r)}),\,r\in\{1,\,\nu\},$ are computed according to \eqref{E:pol_vec};
  \item solve the following convex optimization problem to compute the PCE's coefficients:
\begin{subequations}
\label{E:sparse_optim}
\begin{gather}
\min\limits_{\baa\in \R^{L}}\|\textsc{W}\baa\|_1+\beta\|\tilde{\Lambda}(\tilde{\bv}-\tilde{\bP}\baa)\|_2  \label{opt21a} \\
\text{\emph{subject to}}\nonumber\\
\text{convex constraints}  \label{E:opt_constr}
\end{gather}
\end{subequations}
\end{enumerate}
\end{algorithm}
In \eqref{opt21a}, for a generic vector $\boldsymbol{y}\in\R^m$ the $\ell_1$ and $\ell_2$ vector norms are defined as $\|\boldsymbol{y}\|_1\doteq\sum\limits_{k=1}^{m}|y_k|$ and $\|\boldsymbol{y}\|_2=\left(\sum\limits_{k=1}^{m}y_k^2\right)^{1/2}$, respectively. The diagonal weighting matrix $\textsc{W}$ is defined as $\textsc{W}\doteq\text{diag}\left(\textsc{w}(l_k)\right)\in\R^{L\times L}$, where  $l_k$ is the order of the multi-index $\ba_k$, and $\textsc{w}(l_k),\,k\in\{0,\,L-1\}$ is a sequence of scalar weights with the following properties:
\begin{equation}\label{E:weights}
\begin{array}{rll}
\textsc{w}(l_k)&>&0,\,\,\forall k\in\{0,\,L-1\};\\
\textsc{w}(l_k)&>&\textsc{w}(l_j)\iff l_k>l_j,\,\forall k,j\in\{0,\,L-1\};\\
\max\limits_k\textsc{w}(l_k)&=&1.
\end{array}
\end{equation}
Moreover, $\beta>0$ is a scalar weight. $\textsc{w}(\cdot)$ and $\beta$ are parameters chosen by the user. Finally, the diagonal matrix $\tilde{\Lambda}\doteq\text{diag}\left(\tilde{\lambda}\right)$ contains the values $\tilde{\lambda}\doteq[f_{\bt}(\tilde{\bt}_{(1)}),\ldots,f_{\bt}(\tilde{\bt}_{(\nu)})]^T$ of the
pdf $f_{\bt}$, evaluated at the considered samples $\tilde{\bt}_{(r)},\,r\in\{1,\,\nu\}$.\\
We denote with $\baa^*$ a global minimizer of \eqref{E:sparse_optim} and with $\hat{v}^*(\bt)$ the related PCE. The convex constraints \eqref{E:opt_constr} are optional and they will be better specified later on; we now consider the unconstrained problem \eqref{opt21a} and discuss its features.\\
$\,$\\
\textbf{Weighted $\ell_1$-norm regularization of the coefficients}\\
The convex cost function in \eqref{opt21a} is the weighted sum of two terms. The first one, $\|\textsc{W}\baa\|_1$, is a weighted $\ell_1$-norm of the PCE coefficients, in which the weights increase monotonically with the order of the related multivariate polynomials (see \eqref{E:weights}). $\ell_1$-norm regularization is a well-established technique in function approximation and  regression analysis \cite{Tibshirani96},\cite{Donoho06}, and it is a convex relaxation of the problem of computing an approximation which is \emph{sparsest}, i.e. with minimal number of non-zero terms or, equivalently, with minimal $\ell_0$ quasi-norm, see e.g. \cite{Fuchs05}. However, the use of the weighting matrix $\textsc{W}$ is novel and pertains to the particular properties of polynomial chaos expansions. In practice, minimization of the term $\|\textsc{W}\baa\|_1$ yields an estimated coefficients' vector in which the terms related to higher-order multivariate polynomials have smaller absolute value. The reason for including this term in the cost function \eqref{opt21a} is twofold: on the one hand, it accounts for the fact that, due to the convergence property of polynomial chaos, the absolute values of the PCE coefficients should decrease as the order of the corresponding polynomials in the expansion increases; on the other hand, it avoids over-determination of the fitting problem, when the number $\nu$ of data is lower than the number $L$ of coefficients. Indeed, the $\ell_1$-norm regularization allows one to select an initial overly large maximal order $\overline{l}$, and then to rely on the convex optimization procedure to correctly ``pick'' the terms that have higher relevance, even if the number $\nu$ of sampled data points is much lower than the number of coefficients.\\
We note that, in general, there is no particular reason to believe that the ``best'' (i.e. most accurate) vector of coefficients is actually sparse, that is it has ``few'' non-zero elements (this property holds in some specific cases, see Remark \ref{R:doostan} below). However, there are many contributions in the literature (see e.g. \cite{CaMa47},\cite{XiKa02}-\cite{WaKa05}) showing that the accuracy of the truncated chaos expansion rapidly improves with its order. Therefore, one can expect higher-order terms to be less important, and the related coefficients to be ``small'' in magnitude. Approximations of such expansions should then have the coefficients of higher-order terms that are small in magnitude, or even equal to zero, hence the use of a weighted $\ell_1$ norm in the optimization. The use of the weights $\textsc{W}$ can be seen as prior knowledge that is infused in the estimation process, i.e. the knowledge of the fact that higher-order terms are generally less important.\\
$\,$\\
\textbf{Weighted $\ell_2$-norm fitting of the data}\\
The second term in the cost function \eqref{opt21a} accounts for the fitting error between the sampled values $\tilde{\boldsymbol{v}}$ and the estimate $\tilde{\boldsymbol{\Phi}}\boldsymbol{a}$ given by the PCE. Such a fitting term is weighted by the matrix $\tilde{\Lambda}$. Namely, $\tilde{\Lambda}$ is selected as a $\nu\times\nu$ diagonal matrix, whose diagonal contains the values of the joint pdf $f_{\boldsymbol{\theta}}$ of the input random variables $\boldsymbol{\theta}$, evaluated at the sampled values $\tilde{\boldsymbol{\theta}}_{(r)},r=1,\ldots,\nu$. In this way, the fitting errors $v(\tilde{\boldsymbol{\theta}}_{(r)})-\hat{v}(\tilde{\boldsymbol{\theta}}_{(r)})$ with larger weight in the cost function $\|\textsc{W}\boldsymbol{a}\|_1+\beta\|\tilde{\Lambda}(\tilde{\boldsymbol{v}}-\tilde{\boldsymbol{\Phi}}\boldsymbol{a})\|_2$ are those  pertaining to samples $\tilde{\boldsymbol{\theta}}_{(r)}$, whose values of $f_{\boldsymbol{\theta}}(\tilde{\boldsymbol{\theta}}_{(r)})$ are larger. The obtained solution $\baa^*$ will be such that these fitting errors are smaller than those related to samples $\tilde{\boldsymbol{\theta}}_{(r)}$ with smaller values of
$f_{\boldsymbol{\theta}}(\tilde{\boldsymbol{\theta}}_{(r)})$. The rationale of this choice is to reflect the relative importance of the sampled values according to their pdf, in order to reduce the bias that could be induced by ``low-importance'' samples, i.e. eventual values of $\tilde{\boldsymbol{\theta}}_{(r)}$  whose value of $f_{\boldsymbol{\theta}}(\tilde{\boldsymbol{\theta}}_{(r)})$ is small. Clearly, if all samples have the same importance (e.g. if $f_{\boldsymbol{\theta}}$ is uniform, or if the sampled values $\tilde{\boldsymbol{\theta}}_{(r)}$ have similar values of $f_{\boldsymbol{\theta}}(\tilde{\boldsymbol{\theta}}_{(r)})$), then $\tilde{\Lambda}$ will be close to a scaled identity matrix.\\
We note that if random sampling is used, since the samples $\tilde{\boldsymbol{\theta}}_{(r)}$ are chosen according to the pdf $f_{\boldsymbol{\theta}}$ and since a $\ell_2$-norm fitting criterion is used, as $\nu$ increases the effect of such ``outliers'' is inherently avoided. However, one of the goals of the proposed method is to employ relatively few samples, hence this phenomenon may occur and the described choice of $\tilde{\Lambda}$ greatly improves the obtained performance.\\
Finally, the scalar weight $\beta$ can be used to achieve a tradeoff between the accuracy of the PCE, with respect to the collected data, and its complexity, in terms of  weighted $\ell_1$-norm. In practical applications, with a ``high'' value of $\beta$ the Algorithm \ref{A:SPCE} yields good results and it is quite robust with respect to different choices of weights $\textsc{w}(l)$, provided that the properties \eqref{E:weights} are satisfied (see also Remark \ref{R:choice_nu} below).\\
$\,$\\
\noindent If no constraints are included, problem $\eqref{E:sparse_optim}$ can be cast as a quadratic program, and a global solution can be efficiently computed also with thousands of coefficients \cite{BoVa09}. We now present two kinds of convex constraints \eqref{E:opt_constr}, which can be used to take into account specific additional information on the random variable $v$, at the cost of a possibly higher computational time. These constraints are not meant to be exhaustive of the possibilities that the convex optimization approach can open, when combined with polynomial chaos.\\
$\,$\\
\textbf{Explicit maximal variance constraint}\\
Since the variance $\hat{\s}^2$  of the expansion is a quadratic function of its coefficients (see eq. \eqref{E:PCE_moment_2}), an upper bound $\overline{\sigma}^2\geq0$ on the variance can be explicitly enforced by the convex quadratic constraint:
\begin{equation}\label{E:max_var}
\hat{\s}^2(\baa)\leq\overline{\sigma}^2.
\end{equation}
This constraint is always feasible, since the value $\baa=0$ satisfies it.\\
$\,$\\
\textbf{Explicit bounds on the PCE}\\
If some convex bounds on $v$ are known, e.g. positiveness, these can be easily included in the problem as follows. Assume, without loss of generality, that the bounds can be expressed as $g(v)\leq0$, where $g:\R\rightarrow\R$ is a convex function (multiple convex bounds can be reduced to this form by taking the maximum among all of them). At first, a finite number $\mu$ of further iid samples $\tilde{\bt}_{(r)}$ of $\bt$ has to be computed, together with the corresponding vectors $\P(\tilde{\bt}_{(r)}),\,r\in\{1,\mu\}$ \eqref{E:pol_vec}. Then, the following $\mu$ convex constraints can be included in \eqref{E:opt_constr}:
\begin{equation}\label{E:bounds}
g(\P(\tilde{\bt}_{(r)})\baa)\leq0,\,\forall r\in\{1,\mu\}.
\end{equation}
Indeed, if the computed minimizer $\baa^*$ of \eqref{E:sparse_optim} satisfies the sampled constraints \eqref{E:bounds}, it is not guaranteed that the inequality $f(\hat{v}^*(\bt)))\leq0$ is satisfied \emph{with probability one}; however, some probabilistic results have been established in the context of random convex programming (see \cite{CaCa05,Cala10}), and these can be used to tune the number of constraints $\mu$. As a final remark, we note that, if the bounds on $v$ are such that $g(0)\leq0$, then the convex optimization problem \eqref{E:sparse_optim} is feasible with probability one in the presence of the $\mu$ constraints \eqref{E:bounds}, since these are always satisfied by the value $\baa=0$.

\begin{remark}\label{R:choice_nu}
The proposed approach can be applied no matter how the initial samples $\tilde{\bt}_{(r)},\,i=1,\ldots,\nu$ are selected, however different methods can lead to different results, in terms of number $\nu$ of data required to achieve a good accuracy. We adopt random sampling here, as we found it to be an effective and simple approach. As regards the choice of $\nu$, in principle the higher is the number of data, the better is the obtained accuracy. However, larger values of $\nu$ also imply higher computational cost, to carry out the initial evaluations of $v(\tilde{\bt}_{(r)})$. A simple and  effective way to choose $\nu$ is to start from a low value, to gradually increase the number of data points with which Algorithm \ref{A:SPCE} is carried out, and to employ a stopping criterion in order to assess whether the employed data are sufficient to get good results in terms of approximation error. One possible such stopping criterion is based on the distance, in some norm, between the coefficients computed with two subsequent increasing numbers of data points.  In particular, denoting with $\baa^{*(\nu)}$ the coefficients estimated with a given number $\nu$ of data, one can consider the $\infty$-norm of the difference between two subsequent coefficients' vectors, $\|\baa^{*(\nu+1)}-\baa^{*(\nu)}\|_\infty$. Such distance typically converges quite rapidly to a neighborhood of a fixed value, as $\nu$ reaches some value $\overline{\nu}$. Then, one can take the coefficients computed with these $\overline{\nu}$ data points as the estimate of the PCE's coefficients. It can be observed that the $L_2$-norm of the error $v(\bt)-\hat{v}(\bt)$, between the true process $v(\bt)$ and the values of $\hat{v}(\boldsymbol{\theta})$ computed with the PCE estimated with different values of $\nu$, converges to a small value when $\nu\simeq\overline{\nu}$.  An example of these trends is given in the first application in Section V.\\
A similar approach can be adopted to tune the weighting matrix $\textsc{W}$ and the scalar $\beta$. The main guideline for the choice of the weights in $\textsc{W}$  is given by the properties \eqref{E:weights}: the values of $\textsc{w}(l_k)$ should be strictly increasing with the orders $l_k$ of the corresponding multivariate polynomials. This approach is quite robust with respect to different specific choices satisfying \eqref{E:weights}: in particular, it can be shown that for different values of $\textsc{w}(l_k)$ very similar estimates of the PCE's coefficient can be obtained, by using a sufficiently large value of $\beta$. We provide an example of this behavior in Section V, too.
\end{remark}

\begin{remark}\label{R:l2_reg}
Although we present here the approach using the $\ell_1$-norm regularized cost function \eqref{opt21a}, actually also an $\ell_2$-norm regularization yields similar results, i.e. with a cost function of the following form:
\[
\|\textsc{W}\boldsymbol{a}\|_2^2+\beta\|\tilde{\Lambda}(\tilde{\boldsymbol{v}}-\tilde{\boldsymbol{\Phi}}\boldsymbol{a})\|_2^2.
\]
We note that the weighting matrices $\textsc{W}$ and $\tilde{\Lambda}$ and the scalar $\beta$ have to be chosen according to the same guidelines for both approaches, which are specific to the application of the method to polynomial chaos expansions.
\end{remark}

\begin{remark}\label{R:doostan}
We note that approaches with a similar rationale, i.e. to obtain PCEs from data with low number of nonzero coefficients, has been also proposed in \cite{BlSu10,BlSu11}, by means of an iterative algorithm, and by \cite{DoOw10}, using $\ell_1$-norm or $\ell_0$-norm regularization.\\
The approach of \cite{BlSu11} is significantly different from the one presented here, since the maximal order $\overline{l}$ and the number $\nu$ of data are both gradually increased, until a satisfactory accuracy is reached. Least-squares are used at each iteration to estimate the coefficients, thus the number of data has to be always sufficiently large (2-3 times the number of coefficients, according to \cite{BlSu11}), so that the fitting problem is not over-determined. In our approach, the coefficients are computed in ``one shot'', through the solution of a single convex optimization problem, without any iterative algorithm, and the employed number of data can be very low thanks to the regularization.\\
The work \cite{DoOw10} is more similar to the approach proposed here, since it uses a weighted  $\ell_1$-norm regularization. However, there are several differences between the two methods, including the considered class of problems and initial assumptions, the choice of both weighting matrices $\textsc{W}$ and $\tilde{\Lambda}$, the inclusion of the fitting term $\|\tilde{\Lambda}(\tilde{\boldsymbol{v}}-\tilde{\boldsymbol{\Phi}}\boldsymbol{a})\|_2$ as a constraint instead of a multi-objective optimization. A very important contribution of \cite{DoOw10} is to provide theoretical results about the goodness of the approximation obtained by compressive sensing techniques applied to Polynomial Chaos expansions. To derive similar results in our case appears to be challenging, since we consider less strict assumptions and a wider class of problems. However, the theoretical results of \cite{DoOw10} provide a further justification to our method, in addition to the good performance obtained in the non-trivial examples treated in Section V.\\
Finally neither the methods of \cite{BlSu10,BlSu11} nor the one in \cite{DoOw10} include the possibility to add convex constraints, accounting for further information eventually available on the stochastic process.
\end{remark}

\section{Application examples}\label{S:Example}
In this Section, we present the results obtained by applying the proposed approach in three different fields. In particular, the first example is related to the simulation of an electric circuit, with both parametric uncertainty and a stochastic input. The system has weak nonlinearities, it evolves in continuous time, it has two continuous state variables and 13 input random variables. The second example is concerned with a model for organization innovation \cite{Kopu97}. Such a model is nonlinear, it has seven positive, continuous states and it evolves in discrete time. The number of input random variables is 12. Finally, the third example is in the field of systems biology and presents the evaluation of the effects of extrinsic noise in the simulation of a chemical oscillator. This last model is simulated through the stochastic simulation algorithm (SSA) method \cite{Gill77}, it evolves in continuous time, it has 9 positive, discrete states and 16 input random variables. All together, these examples show how the convex optimization method can be applied in a straightforward way to problems in a broad range of fields and with significant complexity, in terms of number of input random variables, nonlinearities, and constraints on the variables of interest.
\subsection{RLC circuit with stochastic parametric uncertainty and stochastic input}\label{SS:RLC_param}
\begin{figure}
\centerline{
\includegraphics[bbllx=27mm,bblly=143mm,bburx=263mm,bbury=221mm,width=11.00cm,clip]{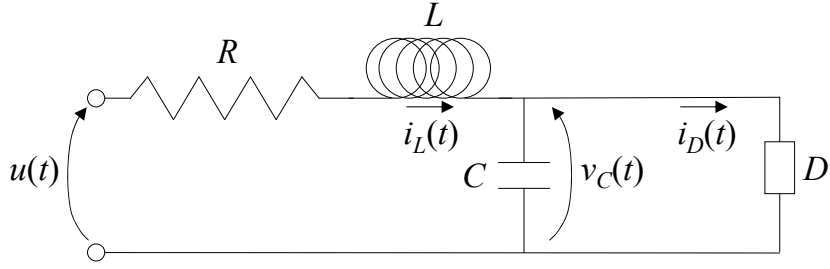}
} \caption{RLC circuit with stochastic parametric uncertainty and stochastic input. Layout of the considered electric circuit.} \label{F:RLC_circuit_2}
\end{figure}

\begin{figure}
\centerline{
\includegraphics[bbllx=5mm,bblly=72mm,bburx=195mm,bbury=215mm,width=10cm,clip]{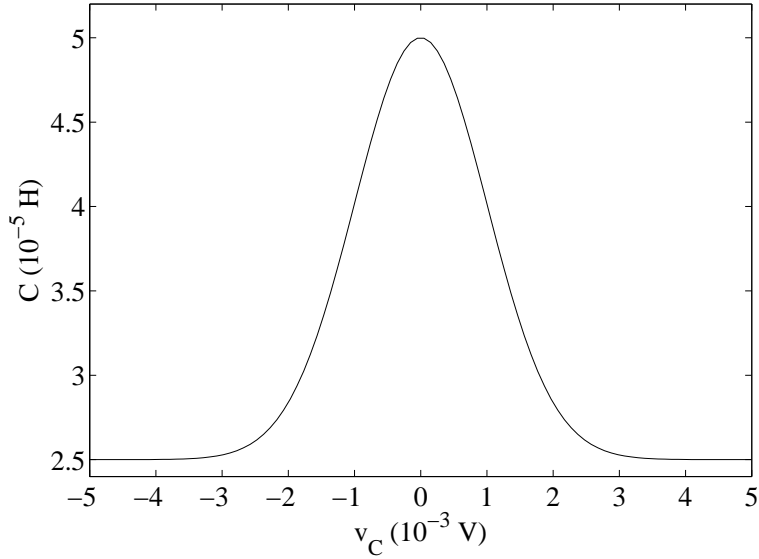}}
\caption{RLC circuit with stochastic parametric uncertainty and stochastic input. Nonlinear characteristic of capacitance $C(v_C)\,(H)$.} \label{F:nonline_circuit_charact}
\end{figure}

\begin{figure}
\centerline{
\includegraphics[bbllx=4mm,bblly=72mm,bburx=200mm,bbury=216mm,width=10cm,clip]{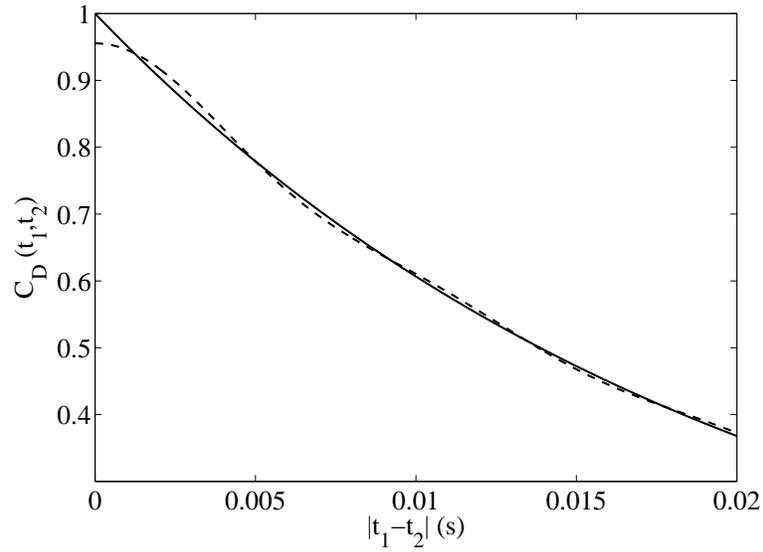}} \caption{RLC circuit with stochastic parametric uncertainty and stochastic input. Comparison between the exact covariance function $C_D(t_1,t_2)$ of the stochastic input (solid line) and the covariance function $\tilde{C}_D(t_1,t_2)$ obtained with the Karhunen-Loeve expansion (dashed line).}\label{F:cov_approx}
\end{figure}

\begin{figure}
\centerline{
\includegraphics[bbllx=-4mm,bblly=75mm,bburx=200mm,bbury=214mm,width=10cm,clip]{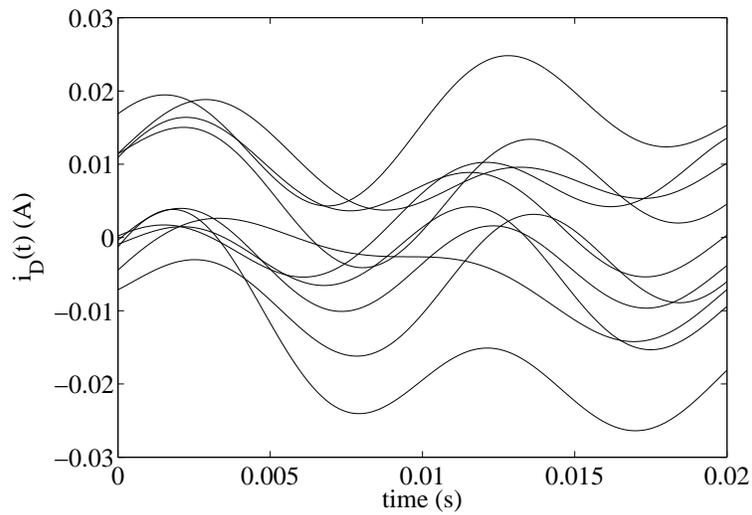}} \caption{RLC circuit with stochastic parametric uncertainty and stochastic input. Examples of 10 different realizations of the stochastic input $i_D(t)$ computed with the Karhunen-Loeve expansion.}\label{F:example_iD}
\end{figure}

\begin{figure}[!hbt]
\centerline{
\includegraphics[bbllx=0mm,bblly=73mm,bburx=200mm,bbury=213mm,width=10cm,clip]{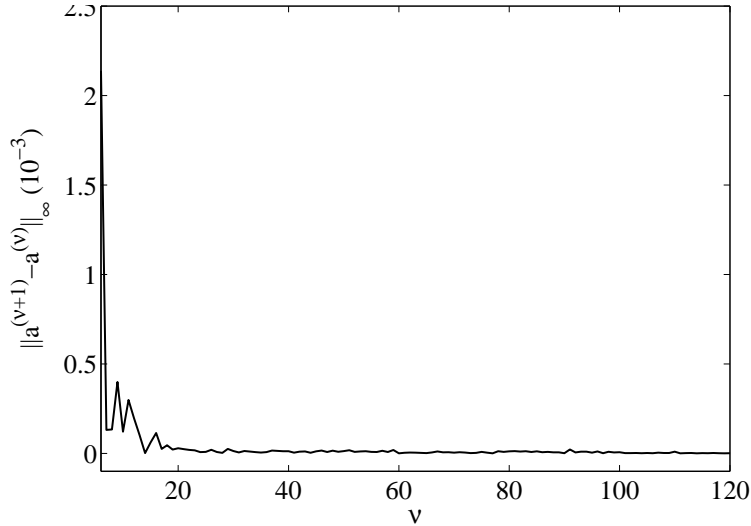}
} \caption{RLC circuit with stochastic parametric uncertainty and stochastic input. Distance $\|\boldsymbol{a}^{*(\nu+1)}-\boldsymbol{a}^{*(\nu)}\|_\infty$ between the expansion's coefficients computed with two subsequent number of data $\nu$, as a function of $\nu$. The plot is related to the voltage at time instant $t_5$, $v_C(t_5,\bt)$.} \label{F:convergence}
\end{figure}

\begin{figure}[!hbt]
\centerline{\begin{tabular}{c}
(a)\\
\includegraphics[bbllx=2mm,bblly=76mm,bburx=195mm,bbury=214mm,width=10.00cm,clip]{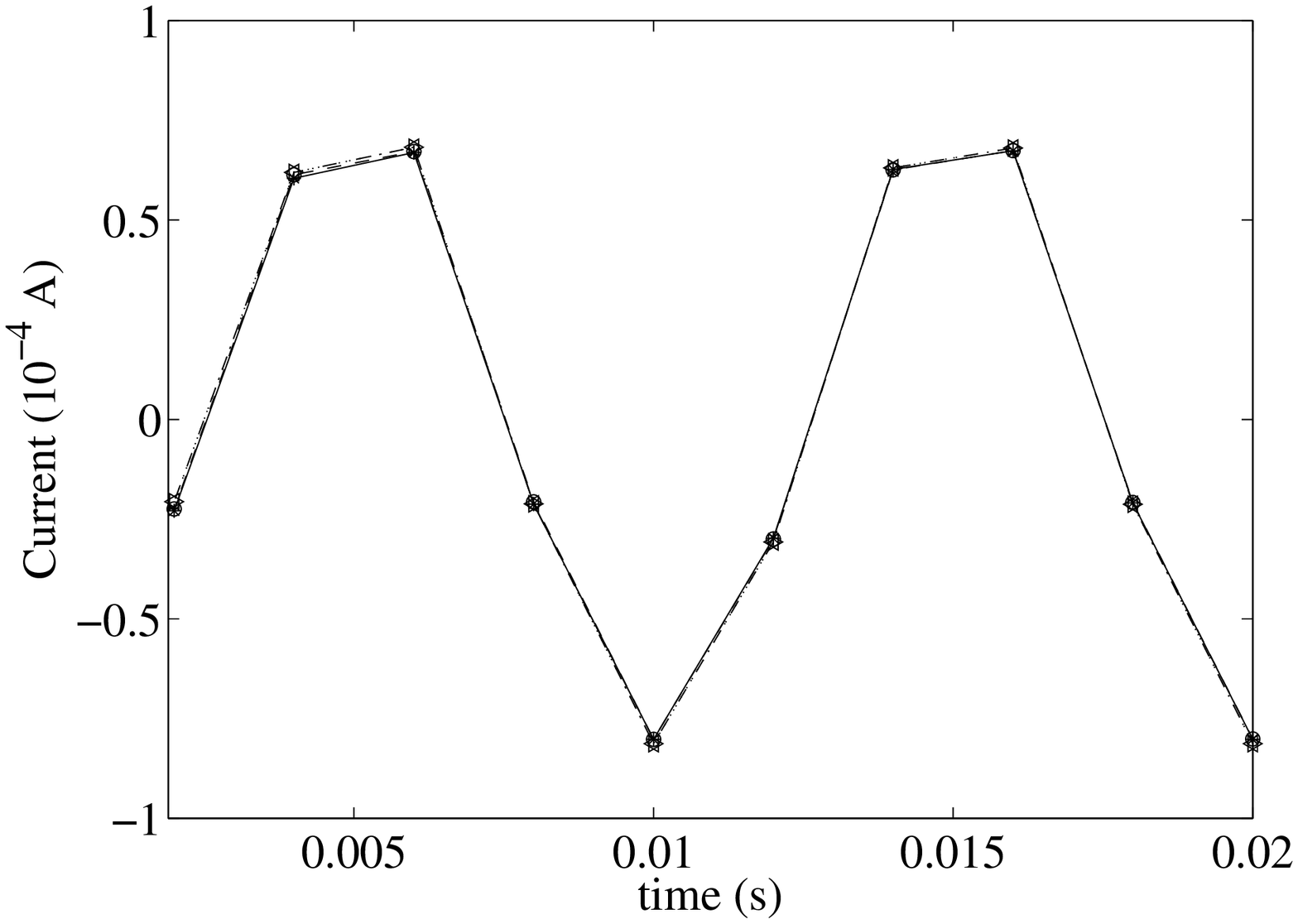}\\
(b)\\
\includegraphics[bbllx=13mm,bblly=77mm,bburx=197mm,bbury=214mm,width=10.00cm,clip]{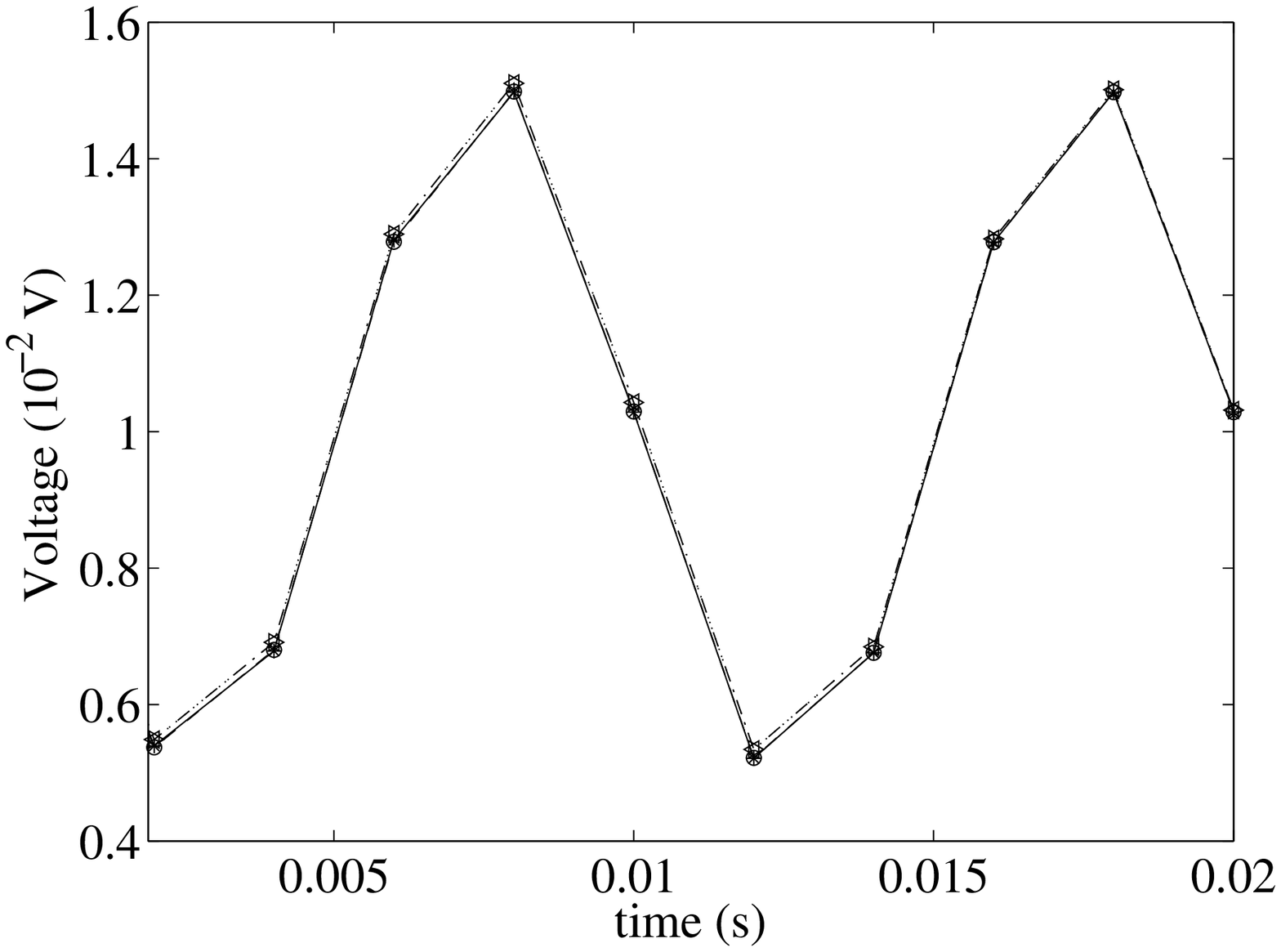}
\end{tabular}}\caption{RLC circuit with stochastic parametric uncertainty and stochastic input. Mean values at $t=t_i,\,i\in\{1,\,10\}$ of (a) current $i_L(t)$ and (b) voltage $v_C(t)$ obtained either with MC simulations of the model (dashed line with '$\circ$') or with the coefficients of the term of degree 0 in the PCEs $\hat{i}_{L}(t_i,\bt),\,\hat{v}_{C}(t_i,\bt)$ (solid line with '$*$') and $\hat{i}_{L}^{LS}(t_i,\bt),\,\hat{v}_{C}^{LS}(t_i,\bt)$ (dotted line with '$\rhd$').} \label{F:mean}
\end{figure}

\begin{figure}[!hbt]
\centerline{\begin{tabular}{c}
(a)\\
\includegraphics[bbllx=13mm,bblly=77mm,bburx=197mm,bbury=214mm,width=10.00cm,clip]{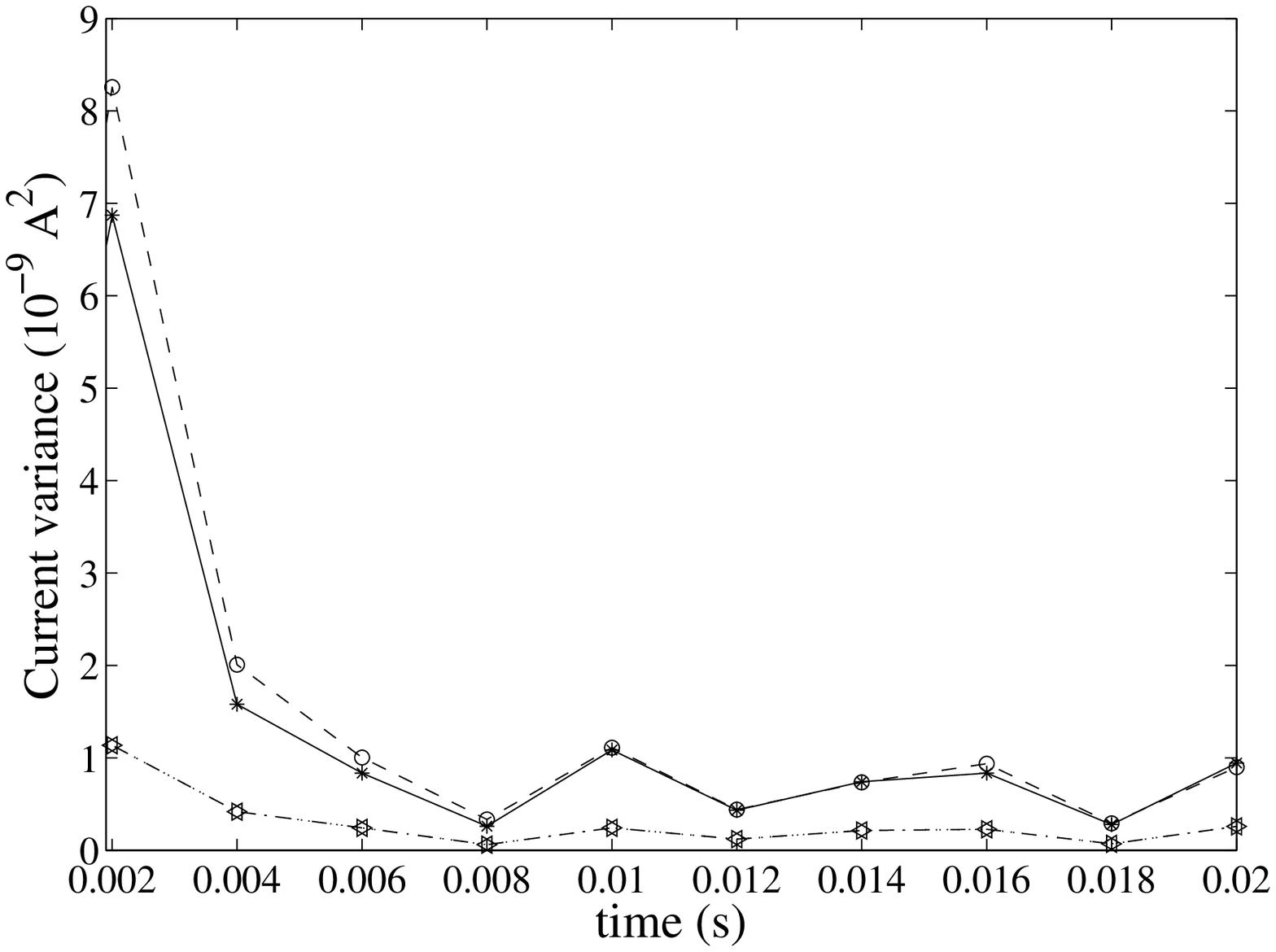}\\
(b)\\
\includegraphics[bbllx=6mm,bblly=83mm,bburx=197mm,bbury=215mm,width=10.00cm,clip]{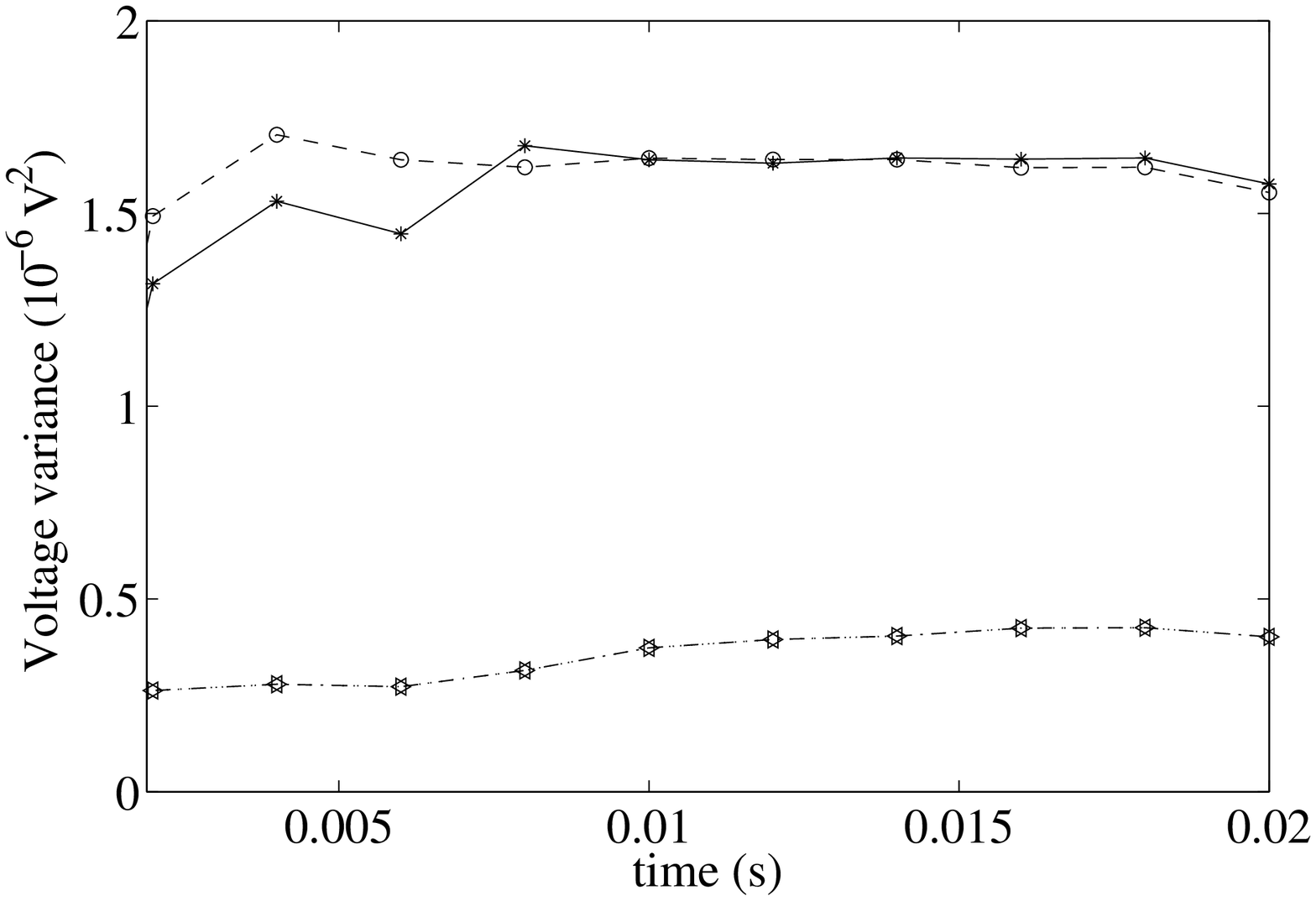}
\end{tabular}}\caption{RLC circuit with stochastic parametric uncertainty and stochastic input. Variances at $t=t_i,\,i\in\{1,\,10\}$ of (a) current $i_L(t)$ and (b) voltage $v_C(t)$ obtained either with MC simulations of the model (dashed line with '$\circ$') or with the coefficients of the PCEs $\hat{i}_{L}(t_i,\bt),\,\hat{v}_{C}(t_i,\bt)$ (solid line with '$*$') and $\hat{i}_{L}^{LS}(t_i,\bt),\,\hat{v}_{C}^{LS}(t_i,\bt)$  (dotted line with '$\rhd$').} \label{F:variance}
\end{figure}

Consider the electric circuit depicted in Fig. \ref{F:RLC_circuit_2}. The system equations are
\begin{equation}\label{E:example_eq}
\begin{array}{lll}
\dot{i}_L(t)&=&-\dfrac{1}{L}v_C(t)-\dfrac{R}{L}i_L(t)+\dfrac{1}{L}u(t)\\
\dot{v}_C(t)&=&\dfrac{1}{C}\left(i_L(t)-i_D(t)\right),
\end{array}
\end{equation}
The resistance $R$ is assumed to be a random variable $R=R_0(1+0.3\,\t_1)$, where $R_0=3.5\,\Omega$ and $\t_1$ is a random variable with uniform distribution over $[-1,1]$. The inductance $L$ and the capacitance $C$ are nonlinear functions of the current  $i_L(t)$  and voltage $v_C(t)$, respectively:
\begin{equation}\label{E:nonlin_charact}
\begin{array}{l}
L(i_L(t))=0.5\,\overline{L}\,(1+\exp(a_1\, i_L(t)^2))\\
C(v_C(t))=0.5\,\overline{C}\,(1+\exp(a_2\, v_C(t)^2)),
\end{array}
\end{equation}
where $a_1=-0.5\,10^{8},\,a_2=-0.5\,10^{6}$. As an example, the function $C(v_C)$ is depicted in Fig. \ref{F:nonline_circuit_charact}.
Moreover, the maximal values $\overline{L}$ and $\overline{C}$, achieved when $v_C(t)=i_L(t)=0$, are equal to $\overline{L}=L_0(1+0.2\,\t_2)$, $\overline{C}=C_0(1+0.2\,\t_3)$, where $\t_2,\,\t_3$ are also random variables with uniform distribution over $[-1,1]$. $\t_1,\,\t_2,\,\t_3$ are assumed to be independent.\\
The circuit is connected to a device $D$, which may act both as load and as generator, by applying a current $i_D(t)$. In particular, $i_D(t)$ is assumed to be a stochastic input of the form:
\[
i_D(t)=a_4\,\sin\left(\dfrac{2\pi}{a_5}\,t\right)+a_3\,i_{D,\text{rand}}(t),
\]
where $a_3=1\,10^{-2}$, $a_4=5\,10^{-3}\,$A, $a_5=1\,10^{-2}\,$s. The term $\sin\left(\dfrac{2\pi}{a_5}\,t\right)$ is a known sinusoidal component, while $i_{D,\text{rand}}(t)$ is a random process with mean $\overline{i}_D=0$ and exponential covariance function $C_D(t_1,t_2)$:
\begin{equation}\label{E:exp_covariance}
C_D(t_1,t_2)=\sigma_D^2\exp^{-\mu_D|t_1-t_2|},
\end{equation}
with $\sigma_D=1$ and $\mu_D=50$. In order to model the random process $i_{D,\text{rand}}(t)$ as a function of a finite number of random input variables, we employ the Karhunen-Loeve (KL) expansion (see e.g. \cite{GaSp91}) with 10 independent random variables $\t_{D,1},\ldots,\t_{D,10}$, uniformly distributed in the interval $[-1,1]$:
\begin{equation}\label{E:KL_expansion}
i_{D,\text{rand}}(t)\simeq\overline{i}_D+\sum\limits_{i=1}^{10}\left(\dfrac{\sqrt{\lambda_{D,i}}}{\sigma_{\t_{D}}}g_{D,i}(t)\t_{D,i}\right).
\end{equation}
In \eqref{E:KL_expansion}, $\sigma_{\t_{D}}=1/\sqrt{3}$ is the standard deviation of the independent random variables $\t_{D,i}$, and $\lambda_{D,i},\,g_{D,i}(t),\,i\in\{1,\,10\}$ are, for a given maximal time range $[-T,T]$, the first ten eigenvalues and eigenfunctions of the integral equation:
\begin{equation}\label{E:int_eq}
\int\limits_{-T}^TC_D(t_1,t_2)g_{D,i}(t_2)dt_2=\lambda_{D,i}g_{D,i}(t_1).
\end{equation}
In the case of the exponential covariance function \eqref{E:exp_covariance}, the eigenvalues $\lambda_{D,i}$ are computed as (see \cite{GaSp91}):
\[
\lambda_{D,i}=\dfrac{2\sigma_D^2\,\mu_D}{\omega_{D,i}^2+\mu_D^2},
\]
where $\omega_{D,i}$ are the solutions to the following transcendental equations:
\[
\begin{array}{l}
\mu_D-\omega_{D,i}\tan(T\,\omega_{D,i})=0,\,i\text{ odd}\\
\omega_{D,i}+\mu_D\tan(T\,\omega_{D,i})=0,\,i\text{ even}.
\end{array}
\]
The corresponding eigenfunctions are:
\[
\begin{array}{l}
g_{D,i}(t)=\dfrac{\cos(\omega_{D,i}\,t)}{\sqrt{T+\dfrac{\sin(2T\,\omega_{D,i}\,t)}{2\,\omega_{D,i}}}},\,i\text{ odd}\\
g_{D,i}(t)=\dfrac{\sin(\omega_{D,i}\,t)}{\sqrt{T-\dfrac{\sin(2T\,\omega_{D,i}\,t)}{2\,\omega_{D,i}}}},\,i\text{ even}\\.
\end{array}
\]
We are interested in simulating the system subject to the constant input $u(t)=1\,10^{-2}\,$V, starting from the steady state conditions $i_L(t_0)=0\,$A, $v_L(t_0)=1\,10^{-2}\,$V, from time $t_0=0$ up to time $t_N=0.02$s. Thus, we choose $T=0.02\,$s to compute the KL expansion. Fig. \ref{F:cov_approx} shows a comparison between the exact covariance function $C_D(t_1,t_2)$ and the covariance function $\tilde{C}_D(t_1,t_2)$ obtained with the KL expansion, which is computed as $\tilde{C}_D(t_1,t_2)=\sum\limits_{i=1}^{10}\left(\lambda_{D,i}g_{D,i}(t_1)g_{D,i}(t_2)\right)$, while Fig. \ref{F:example_iD} shows,
as an example, 10 different realizations of the signal $i_D(t)$. We want to analyze the statistics (first and second order moments, and pdf) of the current $i_L(t_i)$ and voltage $v_C(t_i)$ for $t_i=i\,T_s,\,i\in\{1,\,10\}$, with $T_s=2\,10^{-3}\,$s. The input random variables $\bt$ include the 3 random variables related to parametric uncertainty, $\t_1,\,\t_2,\,\t_3$, plus the 10 random variables involved in the KL expansion, $\t_{D,i},\,i\in\{1,\,10\}$. Thus, there are $n=13$ input random variables in total, all uniformly distributed in the interval $[-1,1]$.
According to Table \ref{T:ortho_polynomials}, the PCE is formulated by using Legendre polynomials, whose coefficients can be easily computed via the recursion \eqref{E:Legendre_recursion}. We applied the convex optimization procedure to estimate the coefficients of a different PCE for the values of $v_C(t_i,\bt)$ and $i_L(t_i,\bt)$ at all the considered time instants. In particular, we carried out $\nu=30$ initial simulations by extracting the corresponding values of $\tilde{\bt}_{(r)},\,r\in\{1,\,\nu\}$ according to its distribution, and we used a maximal order $\overline{l}=2$ for the PCE. This results in $L=105$ multivariate polynomials in the expansion. In particular, the choice of $\nu$ has been carried out by using a procedure like the one described in Remark \ref{R:choice_nu}, i.e. by starting from just $\nu=5$ data points and gradually increasing this number, and evaluating the distance $\|\baa^{*(\nu+1)}-\baa^{*(\nu)}\|_\infty$. As an example, Fig. \ref{F:convergence} shows the obtained result for the voltage $v_C(t_5,\bt)$ at time instant $t_5$. In this example, the same number $\overline{\nu}$ has been used for all the variables of interest, $v_C(t_i,\bt)$ and $i_L(t_i,\bt)\,i=1,\ldots,10$, but in general different values of $\nu$ can be used for each variable.
\begin{figure}[!hbt]
\centerline{
\includegraphics[bbllx=5mm,bblly=76mm,bburx=200mm,bbury=215mm,width=10cm,clip]{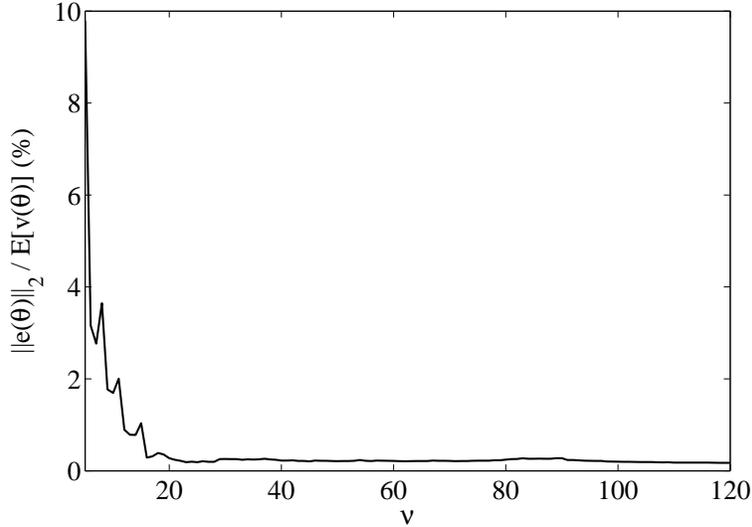}
} \caption{RLC circuit with stochastic parametric uncertainty and stochastic input. Estimate of the $L_2$ norm of the error between the values of $\hat{v}(\boldsymbol{\theta})$, computed with the chaos expansions estimated with different values of $\nu$, and the true process $v(\boldsymbol{\theta})$. The error estimate is expressed as \% of the average $E[v(\boldsymbol{\theta})]$. The plot is related to the voltage at time instant $t_5$, $v_C(t_5,\bt)$.}\label{F:convergence_true}
\end{figure}
It can be clearly noted that there is a number $\overline{\nu}\simeq30$ of points, after which adding new data does not bring significant changes in the coefficients. In Fig. \ref{F:convergence_true} we show the estimate of the $L_2$-norm of the error between the values of $\hat{v}(\bt)$, computed with the chaos expansions estimated with different values of $\nu$, and the true process $v(\bt)$. Such error estimate has been computed through 100,000 Monte Carlo simulations and it is expressed as \% of the average $E[v(\bt)]$ of the true process: we note that such indicator converges rapidly from 10\% to about  0.25\% for  $\nu\simeq\overline{\nu}=30$, and then the increase in number of data does not provide significant improvements. These results justify the use of the difference $\|\baa^{*(\nu+1)}-\baa^{*(\nu)}\|_\infty$ as an indicator to choose the number of samples.\\
Note that the number $\nu=30$ of considered simulations is significantly small as compared to the number of random input variables, $n=13$, and to the number of coefficients that have to be identified, $L=105$. This aspect highlights one of the main advantages of the proposed convex optimization approach, i.e. to be able to obtain quite good accuracy even with an exiguous number of data. Standard methods, like least square fitting, do not share the same feature. In order to carry out a comparison between the approach proposed here and a standard least squares technique, we used the same data $v_C(t_i,\bt_{(r)}),\,i_L(t_i,\bt_{(r)}),\,i\in\{1,\,10\},\,r\in\{1,\,\nu\}$ to identify the PCE coefficients both with our convex optimization procedure and with the following 2-norm minimization problem:
\begin{equation}
\label{E:lq_optim}
\min\limits_{\baa\in \mathbb{R}^{L}}\|\tilde{\bv}-\tilde{\bP}\baa\|_2
\end{equation}
Problem \eqref{E:lq_optim} is a standard least-square regression and it is convex, too, however in this case it is over-determined. Moreover, it does not take into account the available information on the PCE, particularly the fact that the coefficients related to lower-order terms are likely to be more important in the expansion. The PCEs obtained with the convex optimization approach are denoted as $\hat{i}_{L}(t_i,\bt),\,\hat{v}_{C}(t_i,\bt)$, while those obtained by means of least-squares regression are denoted as $\hat{i}_{L}^{LS}(t_i,\bt),\,\hat{v}_{C}^{LS}(t_i,\bt),\,i\in\{1,\,10\}$.\\
In the convex optimization approach, the weights $\textsc{w}(l),\,l\in\{0,\,2\}$ have been chosen as $\textsc{w}(0)=0.00025,\,\textsc{w}(1)=.5,\,\textsc{w}(2)=1$, the scalar weight $\beta$ as $\beta=5$  and the optimization problem has been solved by using the CVX package \cite{cvx} for MatLab$^\circledR$.\\
Fig. \ref{F:mean} shows the courses of
the estimated mean values of $v_C(t_i,\bt)$ and $i_L(t_i,\bt)$ for all of the considered time instants $t_i,\,i\in\{1,\,10\}$, obtained with 100,000 MC simulations and with the PCEs $\hat{i}_{L}(t_i,\bt),\,\hat{v}_{C}(t_i,\bt)$ and $\hat{i}_{L}^{LS}(t_i,\bt),\,\hat{v}_{C}^{LS}(t_i,\bt),\,i\in\{1,\,10\}$. We recall that, for the PCE approximations, the first moment of the process is computed by simply taking, for each $t_i$, the coefficients of the polynomial of degree 0 in the PCEs (see \eqref{E:PCE_moment_1}). It can be noted that both PCE approximations (obtained either with the convex optimization approach proposed here, or with least squares regression) give an accurate estimate of the mean of the variables of interest. However, the results concerning the variance, reported in Fig. \ref{F:variance},
are much different: while the PCE identified with the convex optimization approach proposed here achieves very good results, as compared with the extensive MC simulations, the least squares approach shows a poor accuracy. In particular, the PCEs $\hat{i}_{L}^{LS}(t_i,\bt),\,\hat{v}_{C}^{LS}(t_i,\bt),\,i\in\{1,\,10\}$, identified through the least-square approach, show a much lower variance with respect to the other two estimates. The variances of the PCEs have been computed with the relationship \eqref{E:PCE_moment_2}
\begin{figure}[!hbt]
\centerline{
\includegraphics[bbllx=9mm,bblly=70mm,bburx=194mm,bbury=219mm,width=12.00cm,clip]{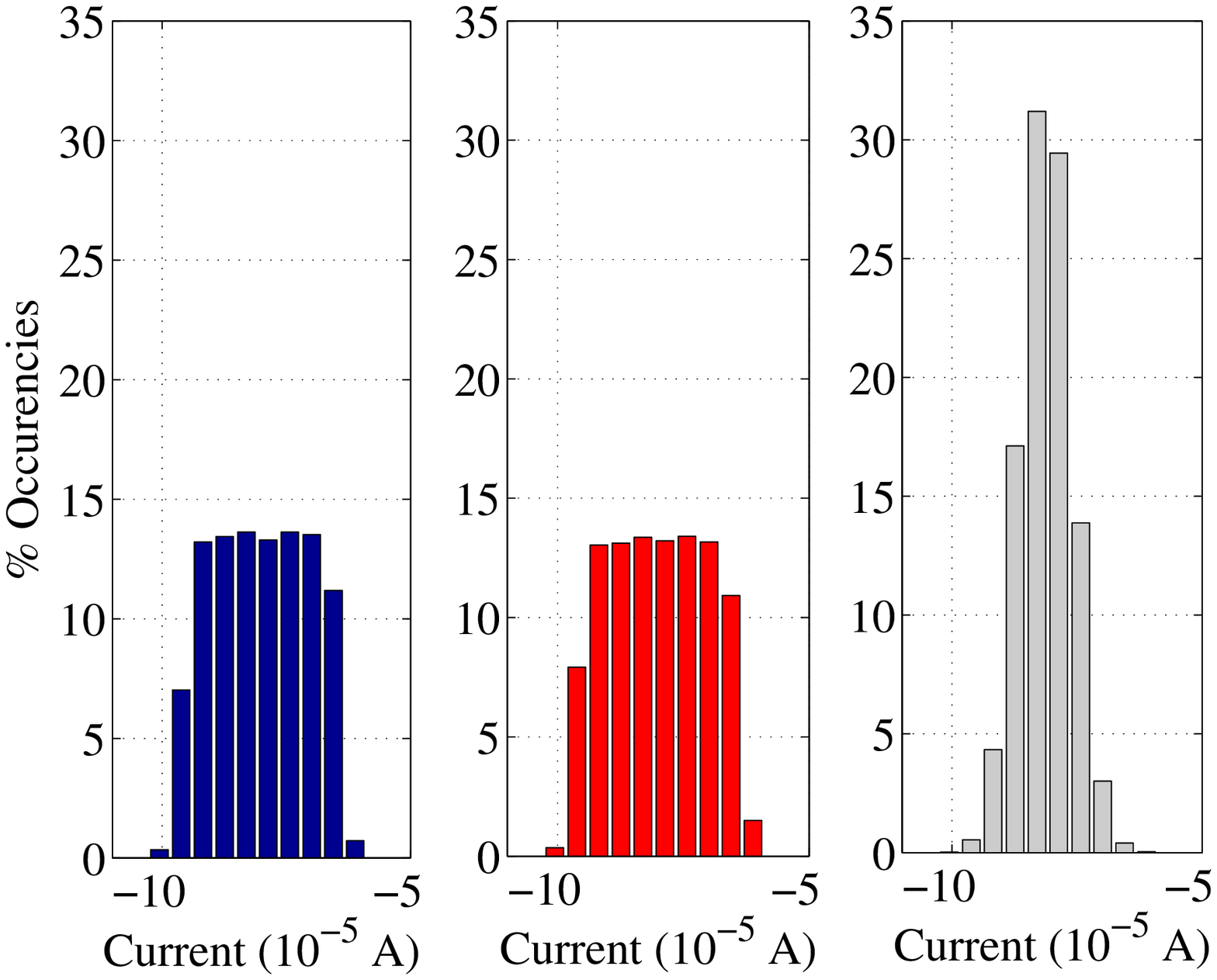}}
\caption{(Color
  online) RLC circuit with stochastic parametric uncertainty and stochastic input.  Comparison between the empirical pdfs of variable $i_L(t_{10},\bt)$ estimated with 100,000 MC simulations with the system model (left) and with 100,000 MC evaluations of the PCEs $\hat{i}_L(t_{10},\bt)$ (middle) and $\hat{i}_{L}^{LS}(t_{10},\bt)$ (right).} \label{F:pdf_es_2}
\end{figure}
(and  \eqref{E:Legendre_norm}, since Legendre polynomials are used here) and they have been compared, in Fig. \ref{F:variance}, to the empirical variance estimated by means of 100,000 MC simulations of the system model.\\
The poor accuracy given by the PCE obtained through least squares regression is further highlighted by the estimates of the pdf, as shown as an example in Fig. \ref{F:pdf_es_2} for the variable $i_L(t_{10},\bt)$: while the empirical pdf computed with the PCE $\hat{i}_L(t_{10},\bt)$ results to be very close to the one computed with the standard MC approach, the one given by  $\hat{i}^{LS}_{L}(t_{10},\bt)$ is very different.\\
In this example, we also show the behavior of the approach with different choices of the weighting matrix $\textsc{W}$ and scalar $\beta$. In particular, we consider the voltage $v_C(t_5,\bt)$ and we estimate its PCE by using weights of the form:
\[
\textsc{W}\doteq\text{diag}\left(\textsc{w}(l_k)\right),\,\textsc{w}(l_0)=1\,10^{-4},\,\textsc{w}(l_k)=\frac{l_k^\zeta}{\overline{l}^\zeta},\,k=1,\ldots,L,
\]
with different values of the exponent $\zeta=1,2,3,4$ (i.e. providing a faster or slower increase of weighting with the order $l_k$). With these different sets of weights, we show the results given by our method, expressed in terms of $\ell_\infty$ distance of the computed coefficients' vectors from the ones presented above, indicated simply as $\boldsymbol{a}^*$, which showed very good accuracy as compared to the actual stochastic process. For each choice of exponent $\zeta$, we also spanned the values of the weighting factor $\beta$, which can be tuned to adjust the relative importance of the regularization term with respect to the fitting term,
\begin{figure}[hbt]
\centerline{
\includegraphics[bbllx=3mm,bblly=74mm,bburx=195mm,bbury=216mm,width=10cm,clip]{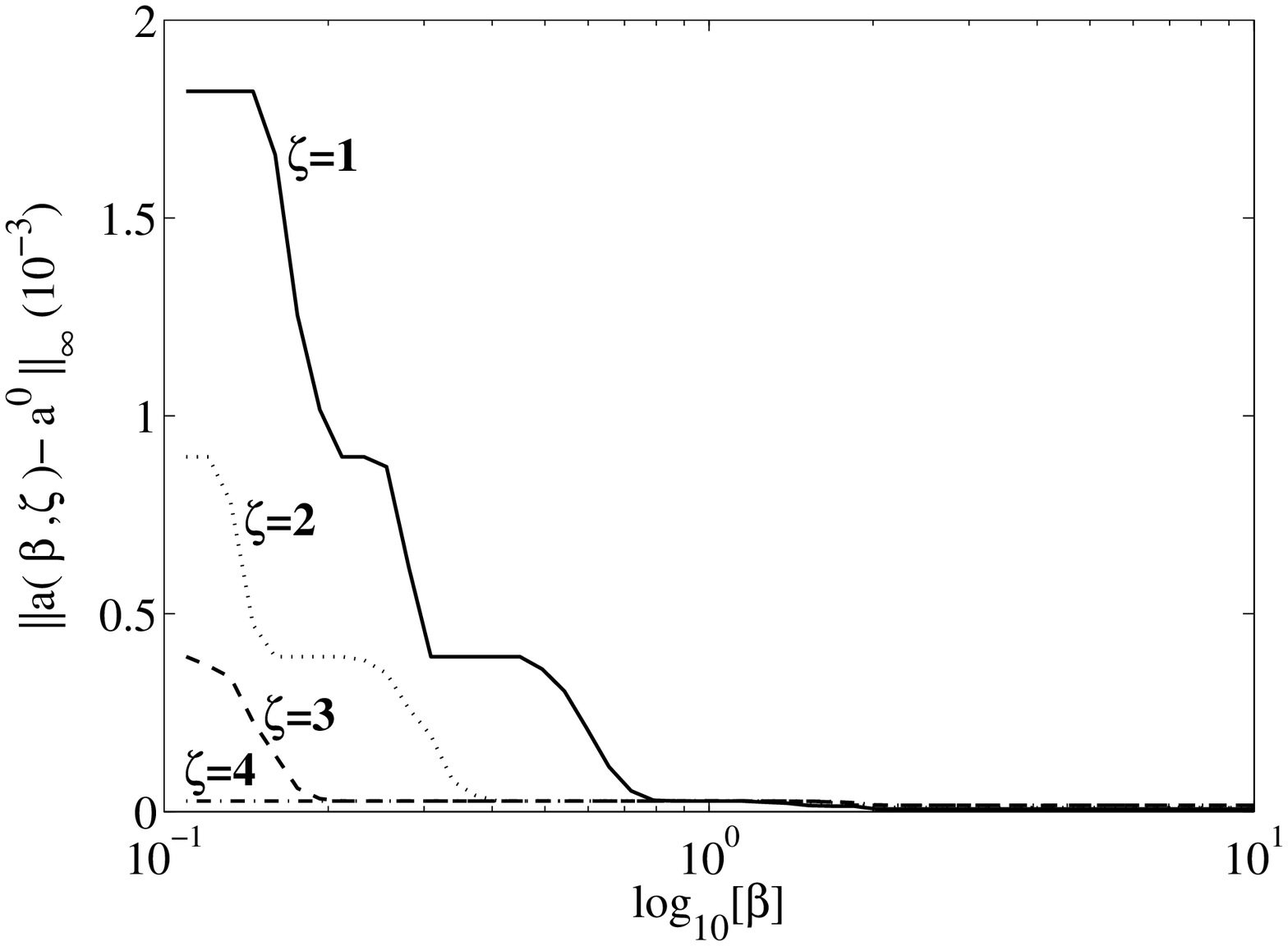}
} \caption{Distance $\|\boldsymbol{a}^*(\beta,\zeta)-\boldsymbol{a}^*\|_\infty$ between the expansion's coefficients computed with different values of the exponent $\zeta$ in the weights $\textsc{w}(l_k)=\frac{l_k^\zeta}{\overline{l}^\zeta}$, and with different values of the scalar weight $\beta$. In particular, the cases $\zeta=1$ (solid line), 2 (dotted), 3 (dashed) and 4 (dash-dot) are shown. The plot is related to the voltage at time instant $t_5$ in the first example of the paper, $v_C(t_5,\bt)$.}\label{F:convergence_weights}
\end{figure}
$\|\tilde{\Lambda}(\tilde{\boldsymbol{v}}-\tilde{\boldsymbol{\Phi}}\boldsymbol{a})\|_2$. Such results, reported in Fig. \ref{F:convergence_weights}, clearly show that convergence to values close to $\boldsymbol{a}^*$ is obtained no matter what kind of weights are chosen (but all satisfying \eqref{E:weights}), by increasing the value of $\beta$. On the other hand, excessively high values of $\beta$ should not be used in order to avoid over-fitting of the data.\\
Finally, as regards the computational times, the 100,000 MC simulations of the model \eqref{E:example_eq} took about 3800$\,$s on a Intel$^\circledR$ Core$^\text{TM}$ 2 Duo processor at 1.3 Ghz, with 4 GB RAM and MatLab$^\circledR$ 2009, while the corresponding 100,000 evaluations of the PCE required about 24$\,$s on the same hardware. The average time required to solve the convex optimization problem in our procedure was 0.45$\,$s, on the same computer.

\subsection{Stochastic model of innovative search}\label{SS:inno_model}
\begin{figure}[!hbt]
\centerline{\begin{tabular}{c}
(a)\\
\includegraphics[bbllx=8mm,bblly=73mm,bburx=195mm,bbury=215mm,width=10.00cm,clip]{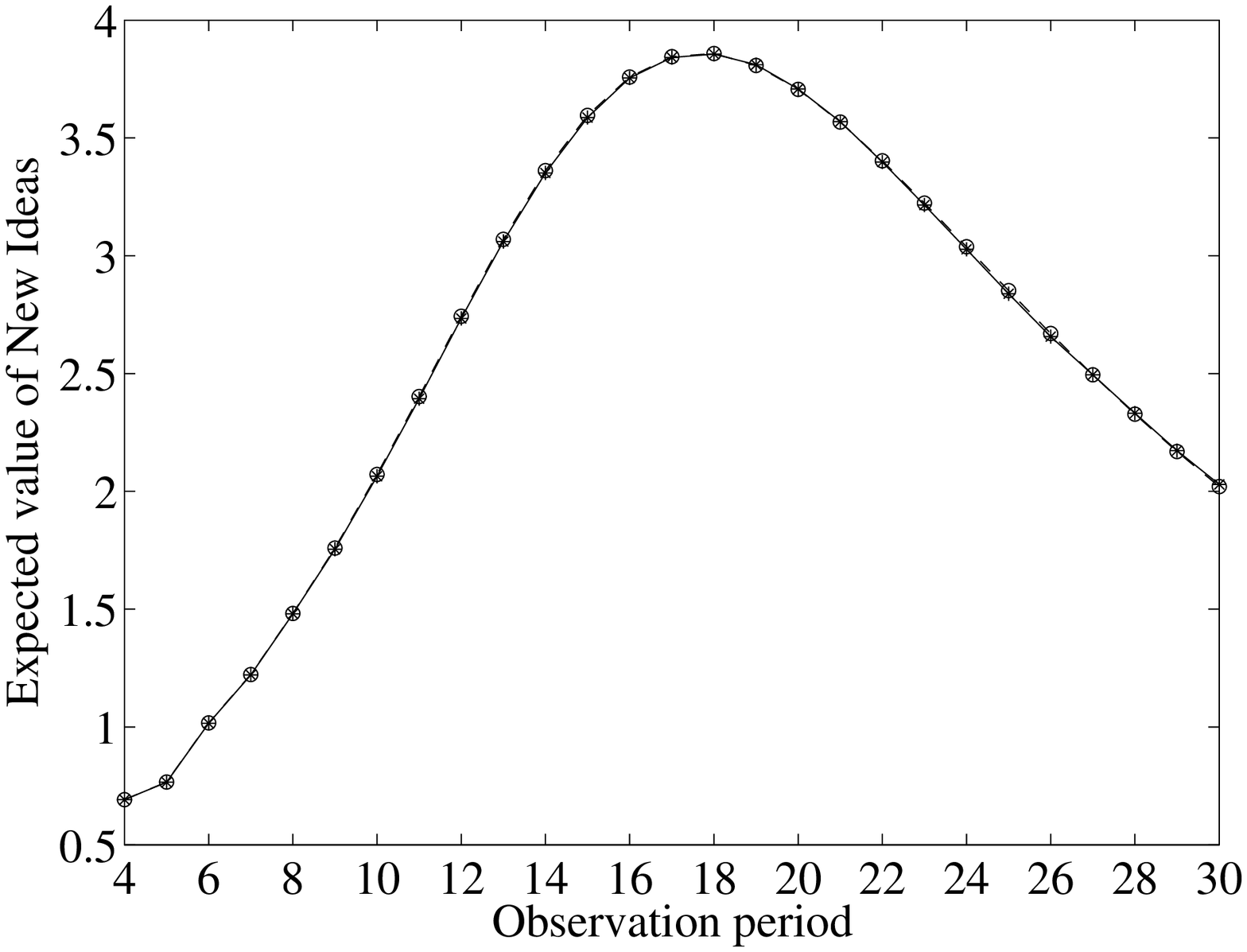}\\
(b)\\
\includegraphics[bbllx=7mm,bblly=72mm,bburx=195mm,bbury=216mm,width=10.00cm,clip]{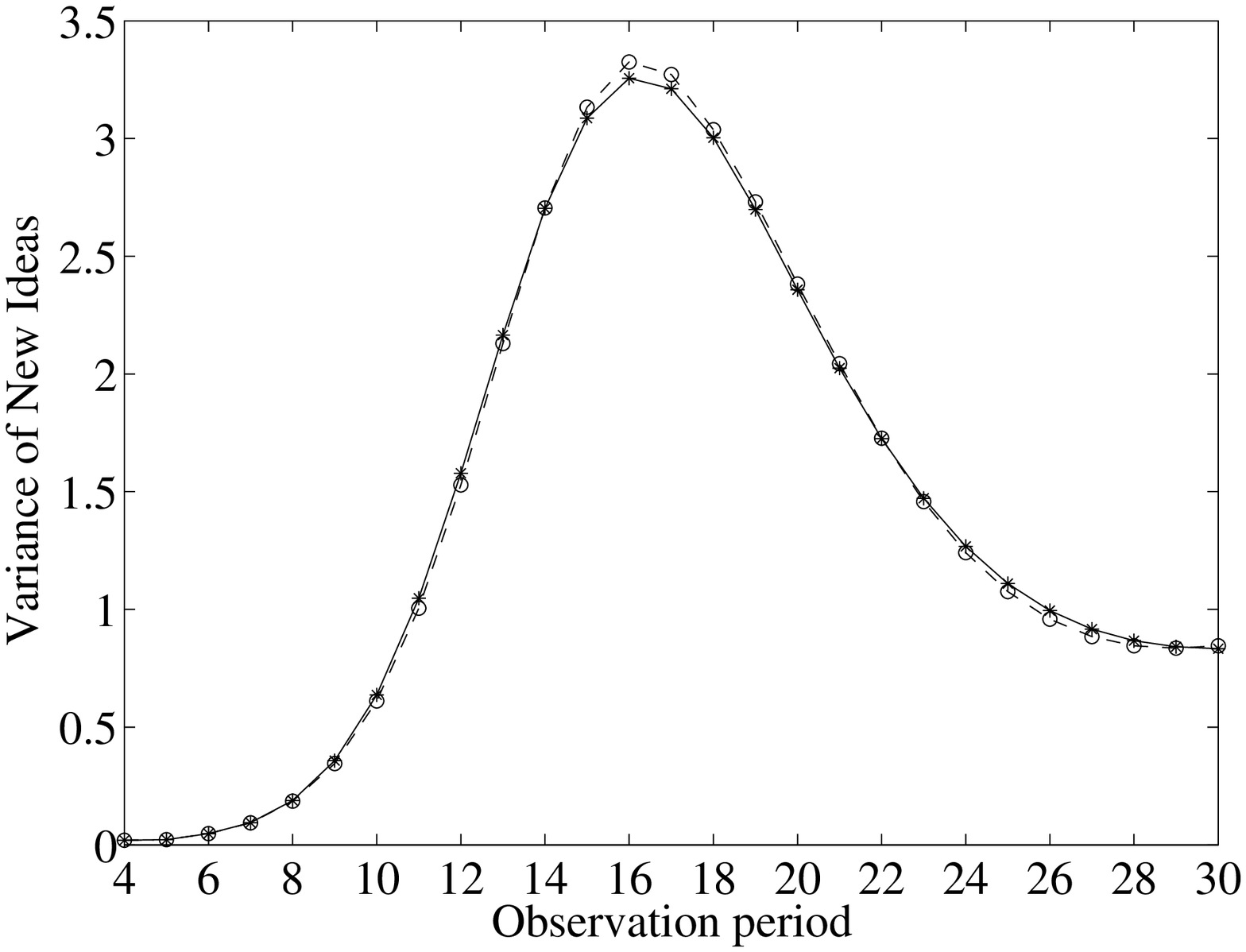}
\end{tabular}}\caption{Stochastic model of innovative search. (a) Mean values and (b) variances of the New Ideas $NI$ at the observation periods $t=4,\ldots,30$, estimated either with 100,000 standard MC simulations (dashed line with '$\circ$') or with the polynomial chaos expansions computed with the convex optimization approach (solid lines with '$*$').} \label{F:inno_mean}
\end{figure}

\begin{figure}[!hbt]
\centerline{\begin{tabular}{c}
(a)\\
\includegraphics[bbllx=9mm,bblly=74mm,bburx=195mm,bbury=216mm,width=10.00cm,clip]{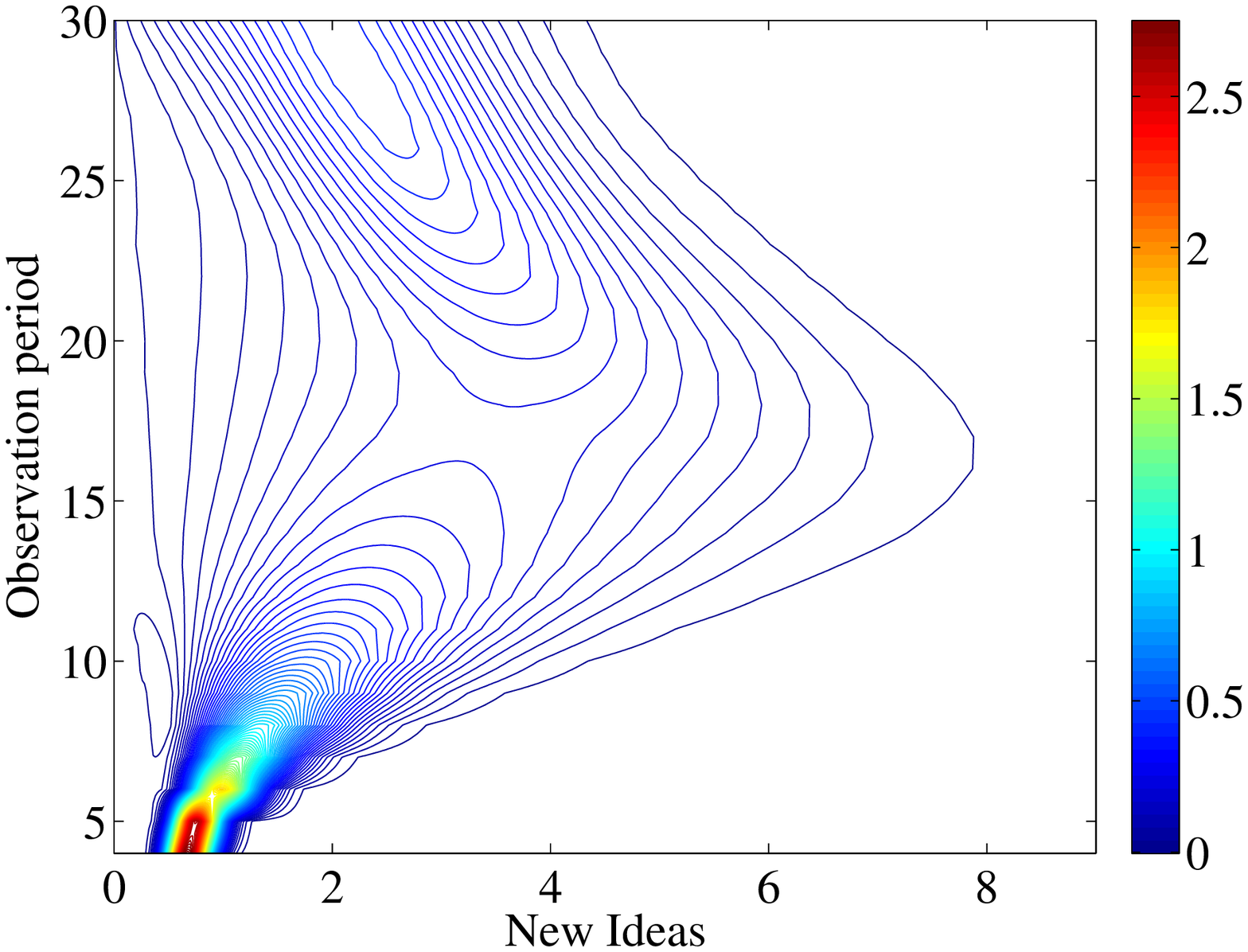}\\
(b)\\
\includegraphics[bbllx=9mm,bblly=74mm,bburx=195mm,bbury=216mm,width=10.00cm,clip]{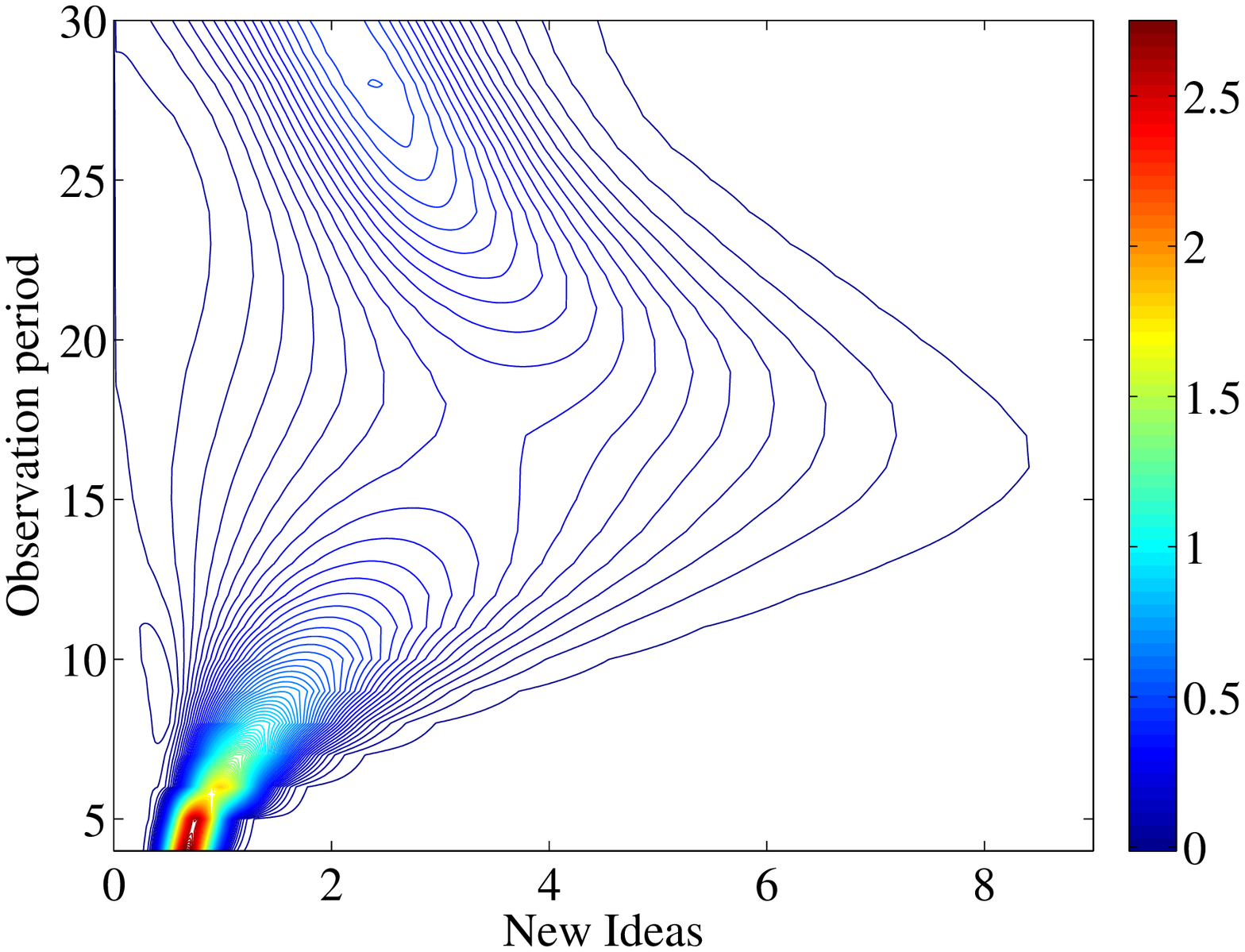}
\end{tabular}}\caption{(Color
  online) Stochastic model of innovative search. Level curves of the pdf of new ideas $NI$ as a function of the observation period, estimated by means of (a) 100,000 standard MC simulations and (b) polynomial chaos expansions computed with the convex optimization approach.} \label{F:inno_flux}
\end{figure}

This second example is concerned with a dynamical model of how organizations pursue innovation, i.e. how they allocate attention to devise new ideas, choose part of them for potential investigation, and finally selects the concepts to be actually tested. The model has been developed by \cite{Kopu97}, it accounts 7 state variables and 12 stochastic parameters, and it has been conceived in discrete time (i.e. the state at the next observation time is a function of the state at the current observation time). An overview of the model equations and of the related parameters is given in Table \ref{T:inno_param}.
\begin{table*}[!hbt]
  \centering
  \caption{States, equations and parameters for the stochastic model of innovative search}\label{T:inno_param}
\begin{tabular}{ll}\hline
\multicolumn{2}{l}{\textbf{State variables}}\\\hline
\multicolumn{2}{l}{Incoming ideas, $II$}\\
\multicolumn{2}{l}{Internal stocks, $IS$}\\
\multicolumn{2}{l}{New ideas, $NI$}\\
\multicolumn{2}{l}{Organizational ideas, $OI$}\\
\multicolumn{2}{l}{Testing ideas, $TI$}\\
\multicolumn{2}{l}{Allocation of attention, $AA$}\\
\multicolumn{2}{l}{External stocks, $ES$}\\\hline
\multicolumn{2}{l}{\textbf{Model equations}}\\\hline
\multicolumn{2}{l}{$II(t+1)=c_1\,ES(t+1)\,AA(t+1)$}\\
\multicolumn{2}{l}{$IS(t+1)=c_2\,IS(t)+ NI(t) + II(t)$}\\
\multicolumn{2}{l}{$NI(t+1)=c_3\,IS(t)$}\\
\multicolumn{2}{l}{$OI(t+1)=c_6\,NI(t+1)+c_4\,IS(t+1)+c_5\,II(t+1)$}\\
\multicolumn{2}{l}{$TI(t+1)=c_7\,TI(t)+ OI(t+1)$}\\
\multicolumn{2}{l}{$AA(t+1)= AA(t)+c_8\,NI(t)+c_9\,NI^2(t)+c_{10}\,TI(t)+c_{11}\,TI^2(t)$}\\
\multicolumn{2}{l}{$ES(t+1)=c_{12}\,ES(t)- II(t)$}\\
\hline
\textbf{Parameter mean} & \textbf{Parameter std. dev.}\\
\hline
$\begin{array}{lll}\overline{c}_1&=&0.1375\end{array}$& $\begin{array}{lll}\sigma_{c_1}&=&0.0225\end{array}$\\
$\begin{array}{lll}\overline{c}_2&=&0.2\end{array}$& $\begin{array}{lll}\sigma_{c_2}&=&0.02\end{array}$\\
$\begin{array}{lll}\overline{c}_3&=&0.5\end{array}$& $\begin{array}{lll}\sigma_{c_3}&=&0.06\end{array}$\\
$\begin{array}{lll}\overline{c}_4&=&0.2\end{array}$& $\begin{array}{lll}\sigma_{c_4}&=&0.02\end{array}$\\
$\begin{array}{lll}\overline{c}_5&=&0.2\end{array}$& $\begin{array}{lll}\sigma_{c_5}&=&0.02\end{array}$\\
$\begin{array}{lll}\overline{c}_6&=&0.5\end{array}$& $\begin{array}{lll}\sigma_{c_6}&=&0.02\end{array}$\\
$\begin{array}{lll}\overline{c}_7&=&0.275\end{array}$& $\begin{array}{lll}\sigma_{c_7}&=&0.025\end{array}$\\
$\begin{array}{lll}\overline{c}_8&=&0.1375\end{array}$& $\begin{array}{lll}\sigma_{c_8}&=&0.0225\end{array}$\\
$\begin{array}{lll}\overline{c}_9&=&-0.0150\end{array}$& $\begin{array}{lll}\sigma_{c_9}&=&0.002\end{array}$\\
$\begin{array}{lll}\overline{c}_{10}&=&-0.0505\end{array}$& $\begin{array}{lll}\sigma_{c_{10}}&=&0.0099\end{array}$\\
$\begin{array}{lll}\overline{c}_{11}&=&0.00055\end{array}$& $\begin{array}{lll}\sigma_{c_{11}}&=&0.00009\end{array}$\\
$\begin{array}{lll}\overline{c}_{12}&=&1.0055\end{array}$& $\begin{array}{lll}\sigma_{c_{12}}&=&0.0009\end{array}$\\
\hline
\end{tabular}
\end{table*}
In particular, the organization allocates a certain quantity of attention, $AA$, to the search for incoming ideas, $II$, from an external stock of ideas and information, $ES$. Part of these incoming ideas is stored in the internal stocks of information, $IS$, from which new ideas $NI$ are derived. The time evolution of the organizational ideas that are actually selected for possible investigation, $OI$, is then influenced by $II$, $IS$ and $NI$. Part of $OI$ feeds the testing ideas, $TI$, i.e. the ideas that, among the organizational ideas, the organization actually chooses to pursue. The state of the system is given by $\boldsymbol{x}=[II,\,IS,\,NI,\,OI,\,TI,\,AA,\,ES]^T$ and it is a vector of non-negative variables (i.e. $\boldsymbol{x}\in\mathbb{R}^{7},\,x_i\geq0\,\forall i\in\{1,\,7\}$). The value of $\boldsymbol{x}$ at the observation period $t+1$ is given by a set of six nonlinear dynamical equations and one algebraic equation, involving the state at the observation period $t$ and 12 parameters $\bc\in\mathbb{R}^{12}$. The parameters are supposed to be independent and distributed according to Gaussian distributions, with standard deviations $\sigma_{c_i}$ and mean values $\overline{c}_i,\,i\in\{1,\,12\}$, as shown in Table \ref{T:inno_param}. Thus, in this example the input random variable $\bt$ is a vector of 12 independent Gaussian variables with normal distribution, such that $c_i=\overline{c}_i+\sigma_{c_i}\t_i,\,i\in\{1,\,12\}$. Extensive MC simulations with this model can be obtained with quite low computational times, however this example is still interesting, from the point of view of the approach proposed in this article, since the model is nonlinear and the considered parameter variations may lead to structural changes in the stability properties of the system (chaotic behavior), from stable modes, to oscillations, to divergent modes. We are interested in computing the time course, over 30 observation periods, of the first and second order moments and of the pdf of new ideas ($NI$), in front of the considered variability in the model parameters. As a matter of fact, for the observation periods $t=1,\,2$ and 3 the value of $NI$ remains fixed at its initial condition, so only the periods $t=4,\ldots,30$ are analyzed. Following our convex optimization procedure, we sample $\nu=300$ values of $\bt\in[-1,1]^{12}$ and compute the corresponding  values of $NI(t,\bt),\,t\in\{4,\,30\}$, starting from the same initial condition $\boldsymbol{x}(0)=[0,\,0\,0,\,0,\,0,\,0.2,\,50]^T$. We use Hermite polynomials (according to Table \ref{T:ortho_polynomials}) and we consider a maximal order $\overline{l}=3$ for the PCEs $\hat{NI}(t,\bt)$, so that each expansion has $L=455$ terms. We note that a Galerkin projection method would lead, in this case, to 3185 discrete-time dynamical equations (i.e. seven model equations, times 455 coefficients in the PCE), while a standard least-squares regression would need at least 455 sampled data, to avoid over-determination. We select the weights $\textsc{w}(l),\,l\in\{0,\,3\}$ as $\textsc{w}(0)=0.0001,\,\textsc{w}(l)=\frac{l^2}{9}\forall l\in\{1,\,3\}$ and the scalar weight $\beta=10^3$. Moreover, since variable $NI$ is defined to be positive, we include 500 additional constraints $\hat{NI}(t,\boldsymbol{\tilde{\t}_{(r)}})\geq0,\,\forall r\in\{1,\,500\},$ as  described in Section IV.
Finally, we include a quadratic constraint on the expansions' variances, by setting $\overline{\sigma}^2=2\,\tilde{\sigma}^2$, where  $\tilde{\sigma}$ is the empiric standard deviation computed by using the 300 simulated data. We solve the convex optimization problem again with the CVX package. The courses of the first and second order moments of $NI(t,\bt)$, computed by using the PCEs' coefficients via \eqref{E:PCE_moment_1}-\eqref{E:PCE_moment_2}, are shown in Fig. \ref{F:inno_mean}(a)-(b), where they are compared with the empirical moments obtained by means of 100,000 MC simulations. Fig. \ref{F:inno_flux} shows the time evolution of the pdf of $NI$, estimated either by computing 100,000 MC simulations or with the corresponding 100,000 evaluations of the PCEs. Finally, Table \ref{T:inno_quartile} shows, as an example,
\begin{table}[!hbt]
  \centering
  \caption{Stochastic model of innovative search: quartiles of the probability distribution of new ideas during 27 observation periods, estimated either with standard MC simulations, or with the PCEs computed with the convex optimization approach}\label{T:inno_quartile}
\begin{tabular}{ccccccc}\hline
\textbf{Obs.}& \multicolumn{3}{c}{\textbf{MC simulations}}&\multicolumn{3}{c}{\textbf{Polynomial chaos}}\\
\textbf{period}& 25\%&50\%& 75\%& 25\%&50\%& 75\%\\\hline
4&  0.59  &  0.68 &   0.78  &  0.59   & 0.68   & 0.78\\
5&    0.67  &  0.77   & 0.87   & 0.66   & 0.76 &   0.86\\
6&   0.86  &  1.00   & 1.16   & 0.87    &1.01&    1.16\\
7&  1.01  &  1.20    &1.41   & 1.00   & 1.19  &  1.41\\
8&   1.17 &   1.43  &  1.74   & 1.17  &  1.44  &  1.74\\
9&    1.32 &   1.68   & 2.10  &  1.33  &  1.68 &   2.10\\
10&    1.50  &  1.96  &  2.52    &1.50   & 1.95 &   2.52\\
11&    1.66   & 2.24  &  2.96  &  1.67  &  2.24&    2.96\\
12&    1.83   & 2.54 &   3.43  &  1.83  &  2.54 &   3.43\\
13&    1.99   & 2.83 &   3.89  &  1.99  &  2.83 &   3.89\\
14&    2.15   & 3.11 &   4.31  &  2.14  &  3.11 &   4.30\\
15&    2.28   & 3.36 &   4.65  &  2.28  &  3.36 &   4.65\\
16&  2.41    &3.55   & 4.89   & 2.40   & 3.56   & 4.89\\
17&    2.51  &  3.70 &   5.01 &   2.52  &  3.71 &   5.01\\
18&    2.58  &  3.77 &   5.00 &   2.60  &  3.78 &   5.01\\
19&    2.63  &  3.77 &   4.90 &   2.65   & 3.78 &   4.92\\
20&  2.65   & 3.70   & 4.72  &  2.66    & 3.70  &  4.73\\
21&    2.62  &  3.57 &   4.49 &   2.65  &  3.59 &   4.51\\
22&    2.56  &  3.41 &   4.24 &   2.58  &  3.41 &   4.23\\
23&    2.46  &  3.22 &   3.99 &   2.50  &  3.23 &   3.98\\
24&    2.33  &  3.04 &   3.75 &   2.39  &  3.05 &   3.73\\
25&    2.18  &  2.85 &   3.52 &   2.25  &  2.86 &   3.48\\
26&    2.03  &  2.66 &   3.30 &   2.11  &  2.69 &   3.26\\
27&    1.87  &  2.49 &   3.11 &   1.98  &  2.53 &   3.09\\
28&    1.70  &  2.31 &   2.93 &   1.81  &  2.36 &   2.90\\
29&    1.54  &  2.15 &   2.76 &   1.65  &  2.21 &   2.75\\
30&    1.38  &  1.99 &   2.61 &   1.49  &  2.05 &   2.60\\
\end{tabular}
\end{table}
a comparison between the quartiles of the variables of interest, computed either with standard MC simulations, or with the corresponding PCE.  The good matching between all of these statistics shows that, also in this case, the expansions' coefficients, computed with the proposed convex optimization method, are able to describe the stochastic process with good accuracy.

\subsection{Chemical oscillator}\label{SS:bio_model}
The last application example is in the field of systems biology. We consider the chemical oscillator analyzed in \cite{VKBL02} and we simulate this system by means of Gillespie's stochastic simulation algorithm \cite{Gill77} (SSA) and the common reaction path (CRP) method proposed in \cite{RaSK10}. In this chemical process, the promoters $P_R$ and $P_A$ control a repressor protein $R$ and an activator protein $A$, respectively. The latter is able to combine with either $P_R$ or $P_A$, giving rise to the complexes $P_{RA}$ and $P_{AA}$, to enhance the transcription of $mRNA_A$ or $mRNA_R$, respectively, with the consequent synthesis of new $A$ or $R$ proteins. The repressor protein $R$ is able to combine with $A$, by forming the intermediate complex $A_R$, and to induce its degradation. The state of this model is given by the quantities of the involved molecules, i.e. $\boldsymbol{x}=[A,\,R,\,P_A,\,P_R,\,P_{AA},\,P_{RA},\,mRNA_A,\,mRNA_R,\,A_R]^T\in\mathbb{R}^9_{\geq0}$, which are discrete by definition, while the model evolves in continuous time. In particular, according to the SSA-CRP simulation method, each one of the 16 reactions that may occur in this process has its own internal clock, and its own stream of random firing times, whose total number, at a given time instant, is a Poisson random variable. We indicate these streams of random firing times as $\xi=\left\{\xi_i\doteq\{\tau_{k_i}\}_{k_i=0}^\infty,\,i\in\{1,\,16\}\right\}$, where $\tau_{k_i}$ is the time interval between two subsequent firing times for the $i^\text{th}$ reaction, and $k_i\in\mathbb{N}$ is a counter giving the total number of reactions of type $i$ that already took place. The internal clocks of the reactions evolve at different speeds, given by the propensities $a_i,\,i=1,\ldots,16$ times the common time variable, $t$. The latter are generally nonlinear functions of the state and of 16 model parameters $c_i,\,i=1,\ldots,16$. The model's chemical reactions, the related parameters and the propensities' equations are described in Table \ref{T:oscillator}.
\begin{table}
  \centering
  \caption{Reactions, propensities and nominal parameter values for the chemical oscillator. The stochastic model parameters are uniformly distributed in the interval $\pm10$\% around the nominal value}\label{T:oscillator}
\begin{tabular}{lll}\hline
\textbf{Reaction} &\textbf{Propensity} & \textbf{Nominal} \\
& &\textbf{parameters}\\\hline
$P_A\xrightarrow{a_1}P_A+mRNA_A$ & $a_1=c_1\,P_A$ &$\overline{c}_1=50$\\
$P_{AA}\xrightarrow{a_2}P_{AA}+mRNA_A$ & $a_2=c_{15}\,c_1\,P_{AA}$ &$\overline{c}_2=0.01$\\
$P_R\xrightarrow{a_3}P_R+mRNA_R$ & $a_3=c_2\,P_R $&$\overline{c}_3=50$\\
$P_{RA}\xrightarrow{a_4}P_{RA}+mRNA_R$ & $a_4=c_{16}\,c_2\,P_{RA}$&$\overline{c}_4=5$\\
$mRNA_A\xrightarrow{a_5}mRNA_A+A$ & $a_5=c_3\,mRNA_A$&$\overline{c}_5=20$\\
$mRNA_R\xrightarrow{a_6}mRNA_R+R$ & $a_6=c_4\,mRNA_R$&$\overline{c}_6=1$\\
$A+R\xrightarrow{a_7}A-R$ & $a_7=c_5\,A\,R$&$\overline{c}_7=50$\\
$P_A+A\xrightarrow{a_8}P_{AA}$ & $a_8=c_6\,P_A\,A$&$\overline{c}_8=1$\\
$P_{AA}\xrightarrow{a_9}P_A+A$ & $a_9=c_7\,P_{AA}$&$\overline{c}_9=100$\\
$P_R+A\xrightarrow{a_{10}}P_{RA}$ & $a_{10}=c_{8}\,P_R\,A$&$\overline{c}_{10}=1$\\
$P_{RA}\xrightarrow{a_{11}}P_R+A$ & $a_{11}=c_{9}\,P_{RA}$&$\overline{c}_{11}=0.2$\\
$A\xrightarrow{a_{12}}\varnothing$ & $a_{12}=c_{10}\,A$&$\overline{c}_{12}=10$\\
$R\xrightarrow{a_{13}}\varnothing$ & $a_{13}=c_{11}\,R$&$\overline{c}_{13}=0.5$\\
$mRNA_A\xrightarrow{a_{14}}\varnothing$ & $a_{14}=c_{12}\,mRNA_A$&$\overline{c}_{14}=1$\\
$mRNA_R\xrightarrow{a_{15}}\varnothing$ & $a_{15}=c_{13}\,mRNA_R$&$\overline{c}_{15}=10$\\
$A_R\xrightarrow{a_{16}}\varnothing$ & $a_{16}=c_{14}\,A_R$&$\overline{c}_{16}=5,000$\\
\end{tabular}
\end{table}
As an example, consider the $4^\text{th}$ reaction and assume it took place already 10 times, i.e. $k_4=10$.  When the $4^\text{th}$ internal clock hits its own next firing time, given by $\sum\limits_{k_4=0}^{11}\tau_{k_4}$, the $4^\text{th}$ reaction takes place again, the counter $k_4$ is augmented by one, and the system state is updated according to the corresponding chemical equation (i.e. the number of $mRNA_R$ molecules is augmented by one), as well as the values of the propensities. Then, the simulation continues with the new propensity values (i.e. the new clocks' ``speeds''). Since each reaction, when it takes place, influences the propensities of the other reactions, the simulation must be carried out in a sequential fashion,  by iteratively computing the reaction that fires next and updating the state, propensities and counters (for more details, the interested reader is referred to e.g. \cite{RaSK10}).  Typically, SSA simulations are carried out for a given initial state and fixed parameters' values, by taking multiple random extractions of the firing times' streams $\xi$ (internal noise), and then some statistic of interest is analyzed. Yet, the model parameters are not fixed and known, rather they can be assumed to be random variables themselves, the so-called extrinsic noise, and it is of interest to study the sensitivity of the SSA outcome to such parameter variations. PCEs have been already applied in the context of systems biology \cite{KiDN07}, by using a projection method and then a quadrature approach to identify the PCE coefficients. As it is also recalled in \cite{KiDN07}, Gauss-Hermite quadrature is efficient up to 3-5 stochastic dimensions. In this example, we analyze the sensitivity to random perturbations in all of the 16 parameters, thus we have 16 stochastic dimensions. Each one of the 16 model parameters $c_i,\,i=1,\ldots,16$, is assumed to be uniformly distributed in the interval $\pm$10\% centered at the corresponding mean value $\overline{c}_i$, as indicated in Table \ref{T:oscillator}. Therefore, the input random variables are given by a vector $\bt=\{\t_i\in[-1,1]\,:\,c_i=(1+0.1\t_i)\overline{c}_i,\,i=1,\ldots,16\}$ of 16 independent variables, each one with uniform distribution in the interval $[-1,1]$. The variables of interest are the expected value $\overline{A}(t,\t)=E[A(t,\t,\xi)]$ and the variance $\sigma_A^2(t,\t)=E[(A(t,\t,\xi)-\overline{A}(t,\t,\xi))^2]$ of the number of molecules of protein $A$, evaluated every 5$\,$s up to $t=50\,$s of simulation time, starting from the initial condition $\boldsymbol{x}(0)=[0,\,177,\,1,\,1,\,0,\,0,\,4,\,0,\,279]^T$. More specifically, we consider the empirical values of $\overline{A}(t,\t),\,\sigma_A^2(t,\t)$, computed by averaging over 1000 SSA simulations.
By using the CPR method \cite{RaSK10}, each SSA realization is associated to its own, fixed seed that generates the random streams $\xi$. In this way, each simulation is evaluated with different values of $\bt$ (i.e. extrinsic noise) but always with the same random stream $\xi$ (i.e. internal noise): therefore, extrinsic noise and internal noise are effectively decoupled, since in practice, for a fixed value of $\bt$, the process is totally deterministic and given by the SSA simulations, each one with its own stream of random firing times. Thus, the only source of randomness lies in the model parameters $\bc\in\R^{16}$. Indeed, the application of Galerkin projection methods appears to be not trivial in this case, due to the particularity of the described SSA method and to the discrete nature of the state variables.
\begin{figure}
\centerline{\begin{tabular}{c}
(a)\\
\includegraphics[bbllx=6mm,bblly=73mm,bburx=195mm,bbury=216mm,width=10.00cm,clip]{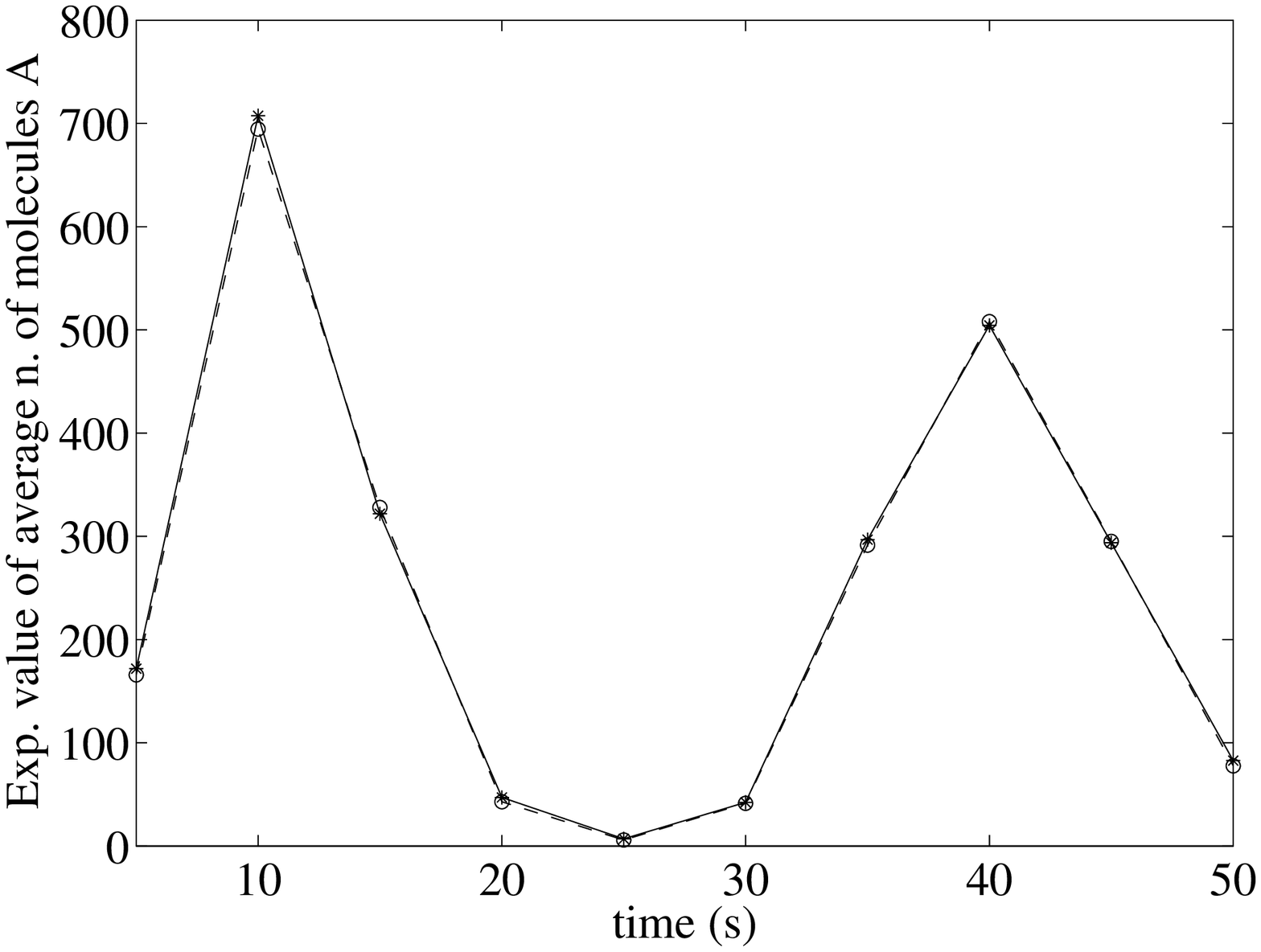}\\
(b)\\
\includegraphics[bbllx=6mm,bblly=74mm,bburx=195mm,bbury=216mm,width=10.00cm,clip]{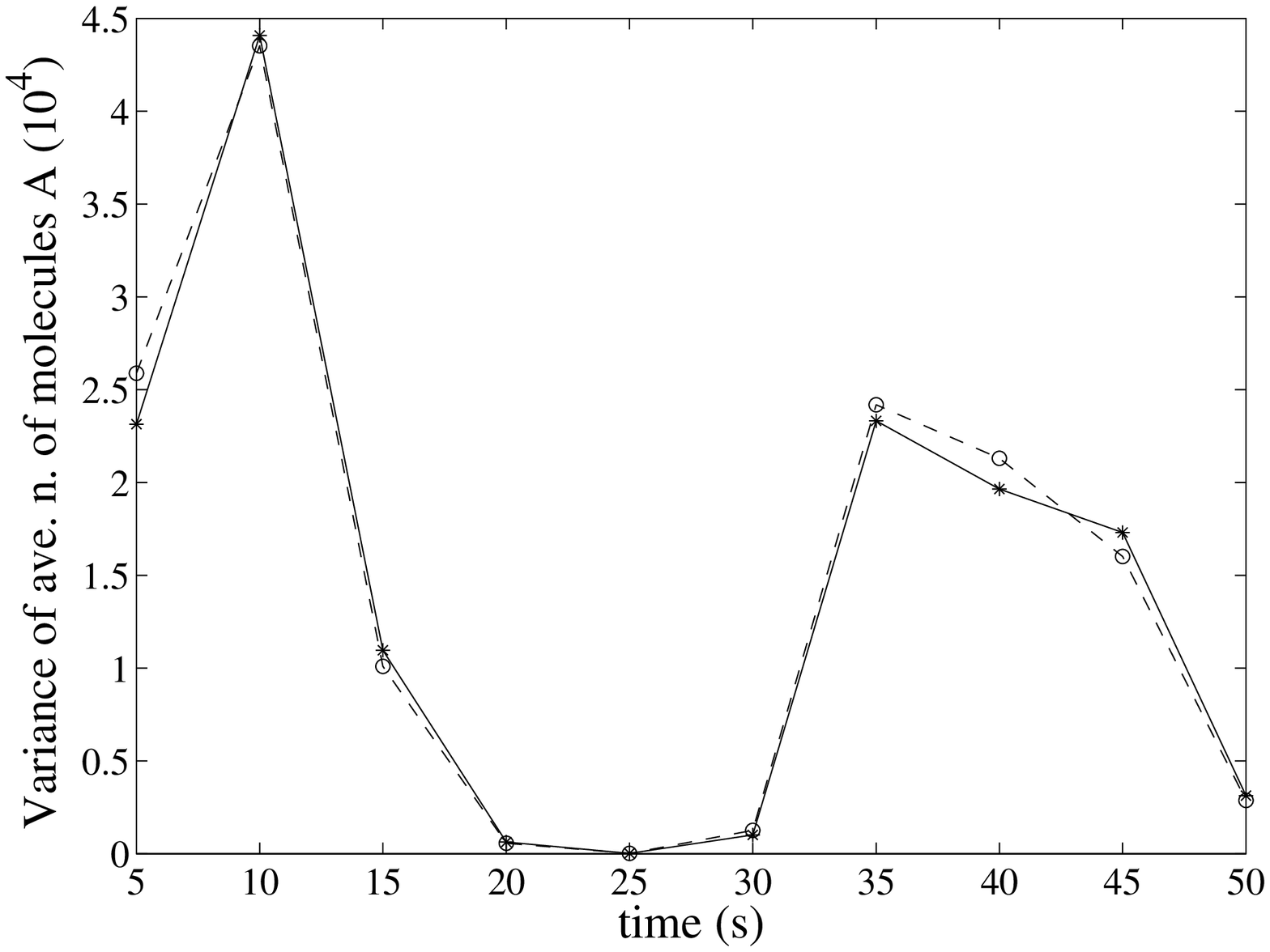}
\end{tabular}}\caption{Chemical oscillator. (a) Mean values and (b) variances of the expected number $\overline{A}(t,\bt)$ of molecules of protein $A$ estimated every 5$\,$s up to 50$\,$s of simulation, either with 10,000 standard MC simulations (dashed line with '$\circ$') or with the polynomial chaos expansions computed with the convex optimization approach (solid lines with '$*$').} \label{F:oscillator_mean}
\end{figure}
\begin{figure}
\centerline{\begin{tabular}{c}
(a)\\
\includegraphics[bbllx=5mm,bblly=70mm,bburx=196mm,bbury=223mm,width=10.00cm,clip]{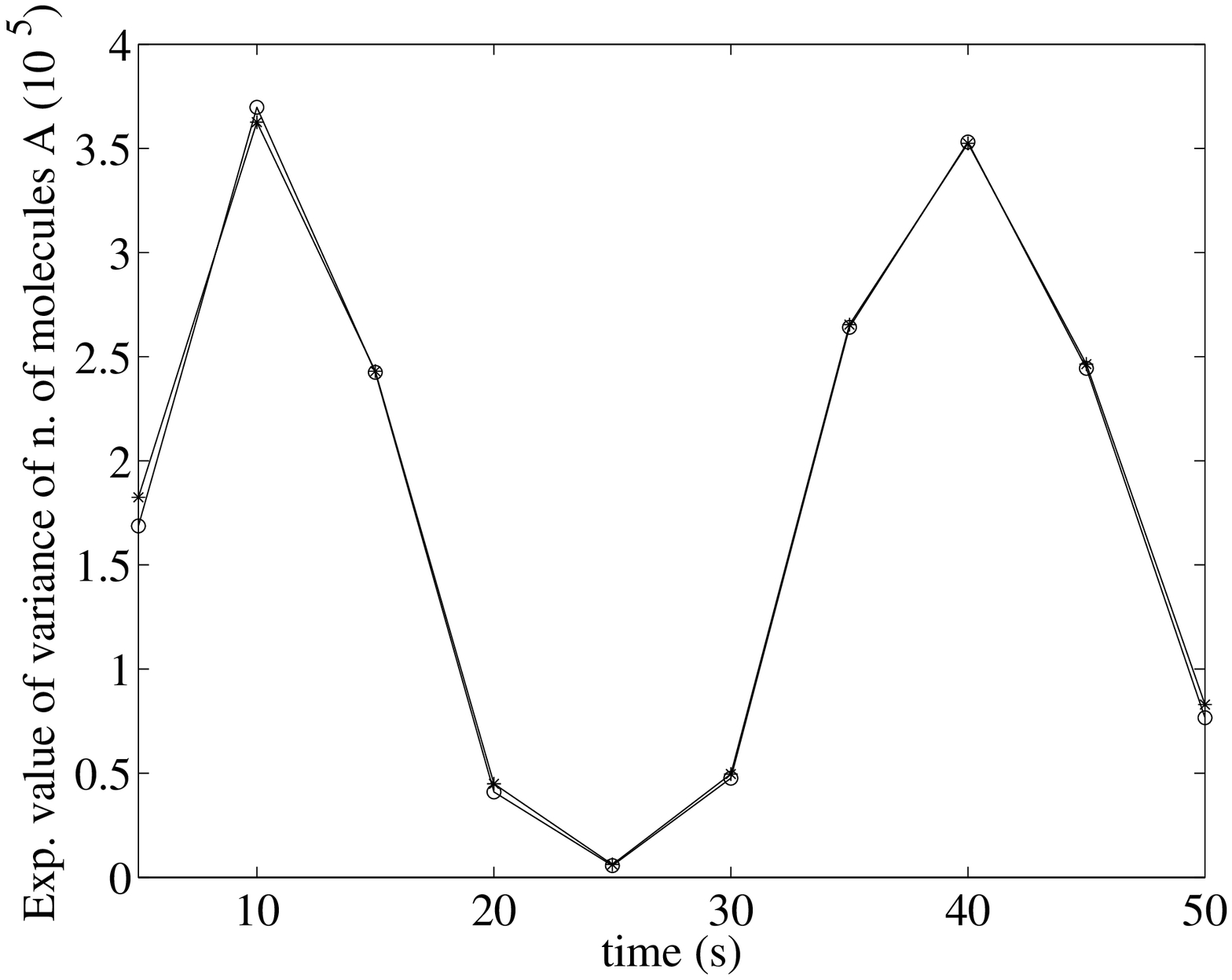}\\
(b)\\
\includegraphics[bbllx=5mm,bblly=70mm,bburx=196mm,bbury=223mm,width=10.00cm,clip]{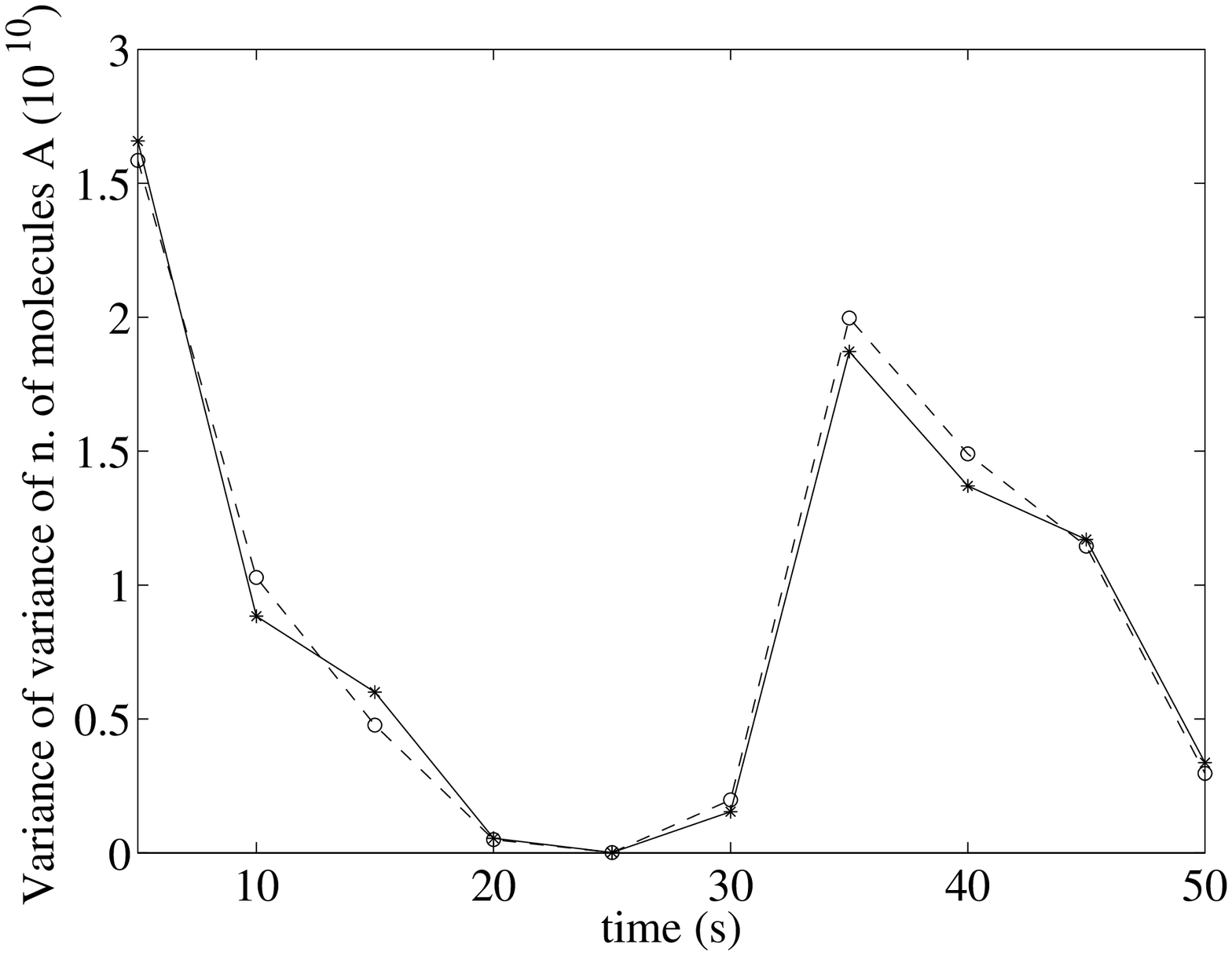}
\end{tabular}}\caption{Chemical oscillator. (a) Mean values and (b) variances of the variance $\sigma_A^2(t,\bt)$ of molecules of protein $A$ estimated every 5$\,$s up to 50$\,$s of simulation, either with 10,000 standard MC simulations (dashed line with '$\circ$') or with the polynomial chaos expansions computed with the convex optimization approach (solid lines with '$*$').} \label{F:oscillator_var}
\end{figure}
\begin{figure}
\centerline{\begin{tabular}{c}
(a)\\
\includegraphics[bbllx=9mm,bblly=73mm,bburx=200mm,bbury=217mm,width=10.00cm,clip]{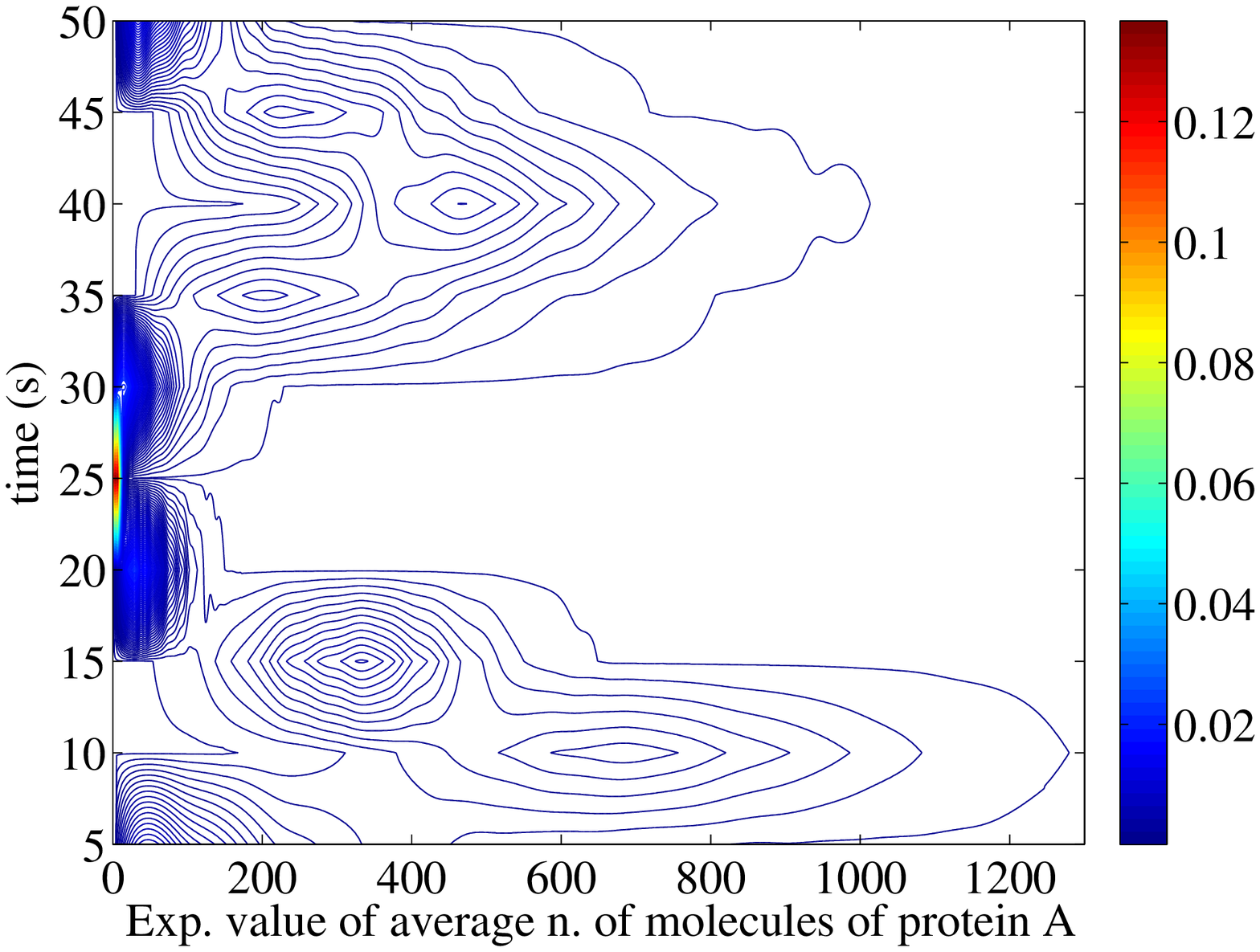}\\
(b)\\
\includegraphics[bbllx=9mm,bblly=73mm,bburx=200mm,bbury=217mm,width=10.00cm,clip]{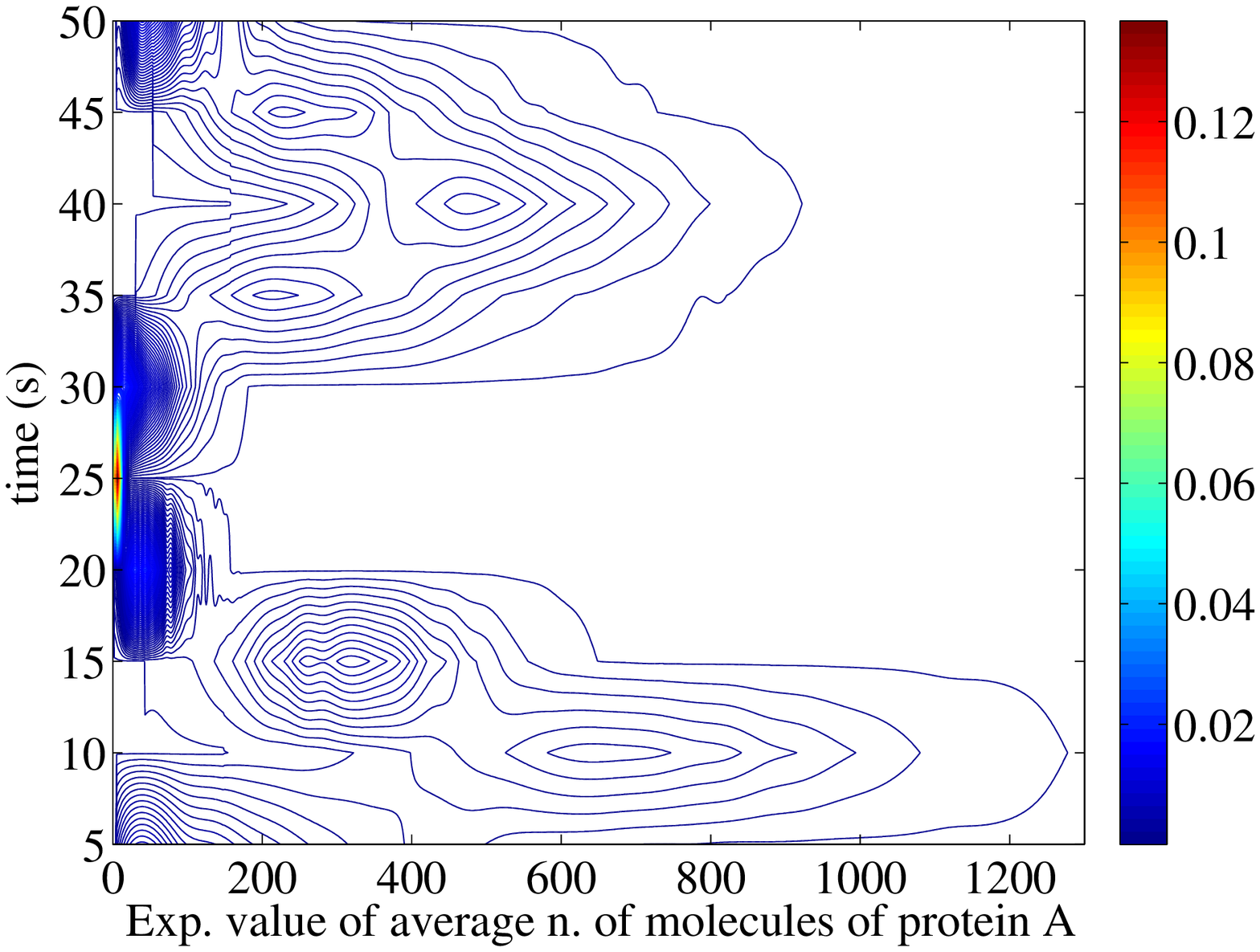}
\end{tabular}}\caption{(Color
  online) Chemical oscillator. Level curves of the pdf of the average value $\overline{A}(t,\bt)$ of the number of protein $A$ molecules as a function of time, estimated by means of (a) 10,000 standard MC simulations and (b) polynomial chaos expansions computed with the convex optimization approach.} \label{F:oscillator_pdf_mean_time}
\end{figure}
\begin{figure}
\centerline{\begin{tabular}{c}
(a)\\
\includegraphics[bbllx=5mm,bblly=69mm,bburx=217mm,bbury=219mm,width=10.00cm,clip]{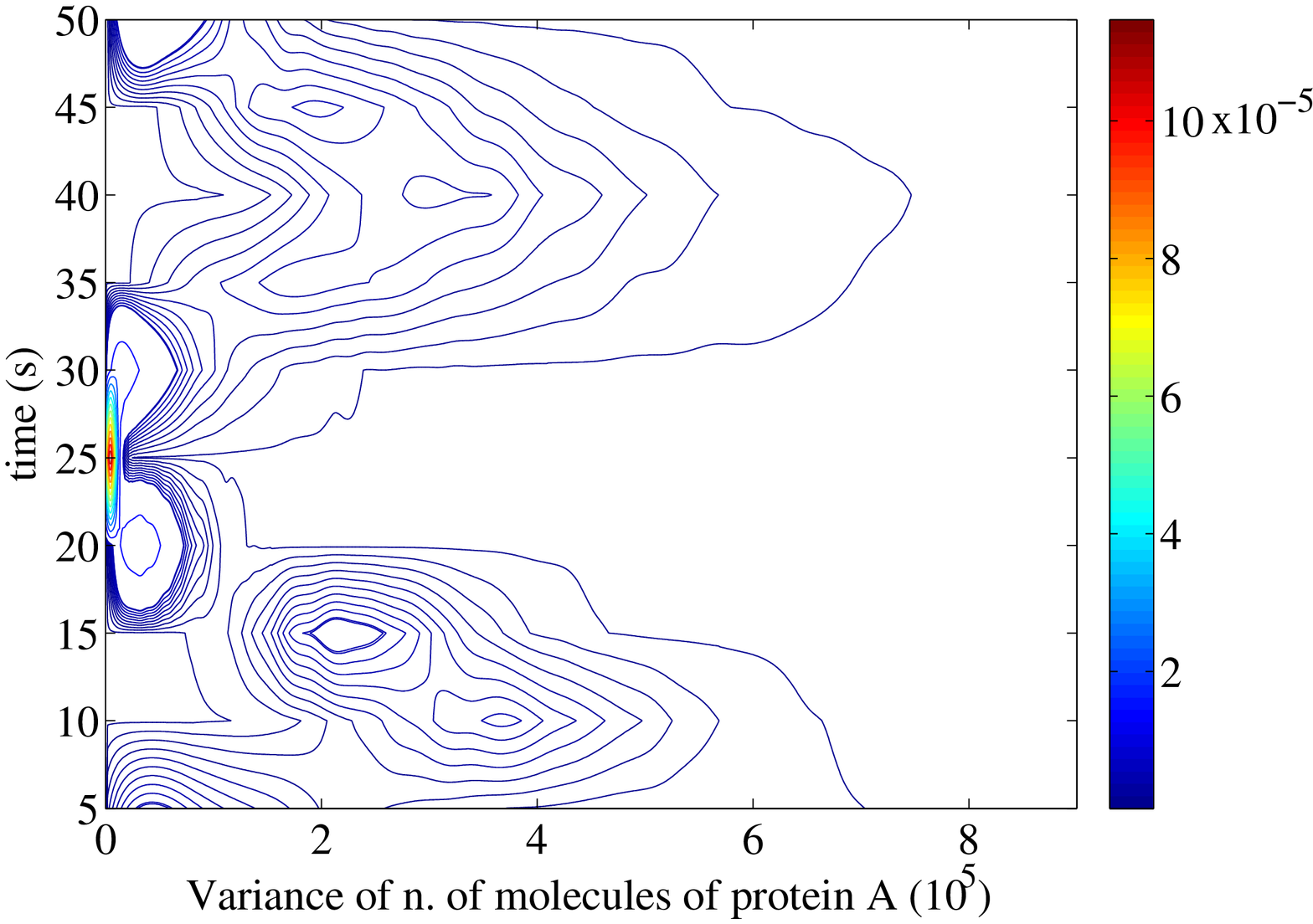}\\
(b)\\
\includegraphics[bbllx=4mm,bblly=69mm,bburx=216mm,bbury=219mm,width=10.00cm,clip]{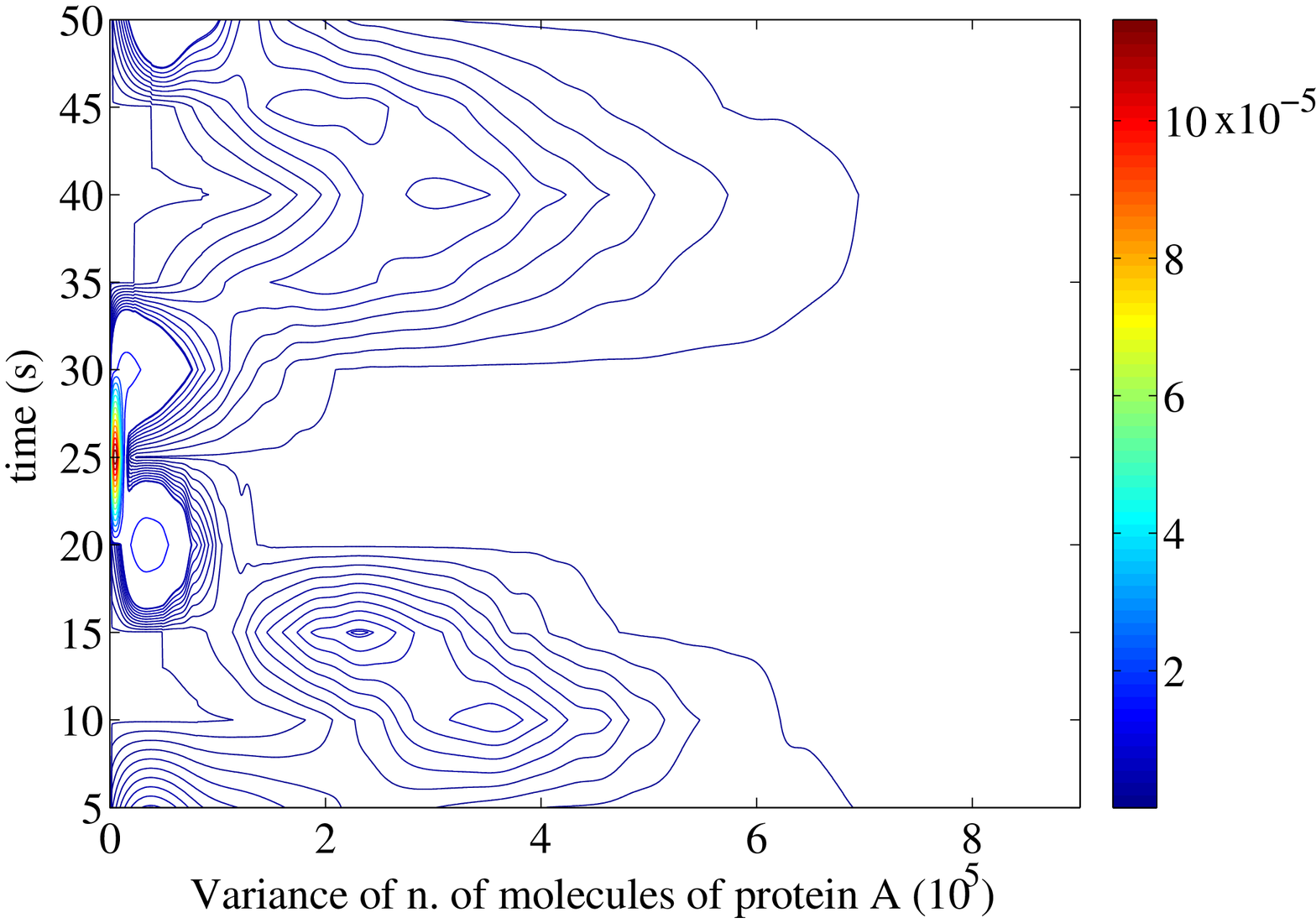}
\end{tabular}}\caption{(Color
  online) Chemical oscillator. Level curves of the pdf of the variance $\sigma_A^2(t,\bt)$ of the number of protein $A$ molecules as a function of time, estimated by means of (a) 10,000 standard MC simulations and (b) polynomial chaos expansions computed with the convex optimization approach.} \label{F:oscillator_pdf_var_time}
\end{figure}
Moreover, we consider PCEs of order 3, which, in a 16-dimensional space, involve 969 terms. With the convex optimization approach, we run 100 sets of 1,000 SSA simulations, corresponding to 100 samples of the random parameter vector. This number of data is very low with respect to the dimensionality of the random parameter $\t\in[-1,1]^{16}$, yet the resulting 3$^\text{rd}$-order PCE is highly accurate with respect to the results obtained with a standard MC approach, considering 10,000 sets of 1,000 SSA simulations. In particular, the comparison between the two approaches is shown in Fig. \ref{F:oscillator_mean}-\ref{F:oscillator_pdf_var_time} and in Tables \ref{T:oscillator_quartiles}-\ref{T:oscillator_quartiles_var}.
\begin{table}
  \centering
  \caption{Chemical oscillator: quartiles of the probability distribution of the expected number $\overline{A}(t,\bt)$ of molecules of protein $A$, computed every 5$\,$s up to 50$\,$s, estimated either with standard MC simulations, or with the PCEs computed with the convex optimization approach}\label{T:oscillator_quartiles}
\begin{tabular}{ccccccc}\hline
 & \multicolumn{3}{c}{MC simulations}&\multicolumn{3}{c}{Polynomial chaos}\\
\textbf{Time (s)}& \textbf{25\%}&\textbf{50\%}& \textbf{75\%}& \textbf{25\%}&\textbf{50\%}& \textbf{75\%}\\\hline
5&     58  &  113 &  217 &   51 &  123 &  242\\
10&   547   & 686 &  833 &  556 &  693 &  841\\
15&   258   & 325 &  392 &  251 &  319 &  388\\
20&    25   &  38 &   56 &   29 &   44 &   61\\
25&     3  &   5  &   8  &   4  &   7  &   9\\
30&    16  &  30  &  54  &  17  &  33  &  58\\
35&   174  & 263  & 381  & 181  & 272  & 386\\
40&   403  & 489  & 591  & 404  & 493  & 595\\
45&   200  & 280  & 371  & 196  & 278  & 373\\
50&    37  &  63  & 104  &  42  &  67  & 110\\
\end{tabular}
\end{table}
\begin{table}
  \centering
  \caption{Chemical oscillator: quartiles of the probability distribution of the variance $\sigma_A^2(t,\bt)$ of the number of molecules of protein $A$, computed every 5$\,$s up to 50$\,$s, estimated either with standard MC simulations, or with the PCEs computed with the convex optimization approach}\label{T:oscillator_quartiles_var}
\begin{tabular}{ccccccc}\hline
 & \multicolumn{3}{c}{MC simulations ($\times10^3$)}&\multicolumn{3}{c}{Polynomial chaos ($\times10^3$)}\\
\textbf{Time (s)}& \textbf{25\%}&\textbf{50\%}& \textbf{75\%}& \textbf{25\%}&\textbf{50\%}& \textbf{75\%}\\\hline
 5&      54.4    &  115.2    &  229.3   &    53.7   &   129.2   &   259.5\\
 10&     298.5   &   364.7   &   436.4  &    294.4  &    356.5  &    422.6\\
 15&     192.7   &   235.6   &   285.8  &    187.2  &    238.5  &    295.9\\
 20&      24.3   &    37.4   &    53.5  &     28.4  &    42.0   &    58.7\\
 25&       3.0   &     5.2   &     7.8  &      3.7  &      5.7  &      8.1\\
 30&      17.1   &    32.9   &    62.3  &     19.3  &     38.1  &     68.9\\
 35&     159.3   &   238.8   &   346.0  &    161.1  &    242.5  &    344.2\\
 40&     264.3   &   335.9   &   424.5  &    264.1  &    335.3  &    422.1\\
 45&     164.7   &   226.4   &   305.0  &   164.5   &   231.1   &   311.2\\
 50&      35.6   &    61.1   &   103.2  &     41.4  &     66.4  &    109.2\\\hline
\end{tabular}
\end{table}
In the convex optimization problem, we chose the weights $\textsc{w}(l),\,l\in\{0,\,3\}$ as $\textsc{w}(0)=0.0001,\,\textsc{w}(l)=\frac{l^3}{27}\forall l\in\{1,\,3\}$, and the scalar weight $\beta=10^3$. We solved the convex optimization problem by using the Yalmip \cite{yalmip} toolbox for MatLab$^\circledR$. Moreover, similarly to the second application example, we included 5,000 additional constraints $\hat{\overline{A}}(t,\boldsymbol{\tilde{\t}_{(r)}}),\hat{\sigma}_A^2(t,\boldsymbol{\tilde{\t}_{(r)}})\geq0,\,\forall r\in\{1,\,5,000\}$, in order to take into account the positiveness of the variables of interest. As regards the computational times, the time required to compute the 100 data points used to identify the PCEs' coefficient was 67 min, the solution of the 20 convex optimization problems (2 variables of interest evaluated at 10 different time instants) took 2 hours (averagely 6 minutes for each PCE), finally the evaluation of 10,000 MC values of the resulting PCE took 8$\,$s on an Intel$^\circledR$ Core$^\text{TM}$ 2 Duo processor at 1.3 GHz, with 4 GB RAM and MatLab$^\circledR$ 2009. Thus, the PCE-convex optimization method took about 3 hours in total. The time required to compute the 10000 standard MC simulations was about 6670 minutes, i.e. 4.6 days. The model equations for the SSA simulations have been programmed in Simulink$^\circledR$, and the computation have been carried out on a Speedgoat$^\circledR$ real-time machine, by using Embedded Matlab$^\circledR$ and xPC-target$^\circledR$ tools to automatically generate the simulation code from the Simulink model. Indeed, the results of this example confirm that the proposed method, based on convex optimization, is able to compute the PCE's coefficients with good accuracy also in the presence of a relatively large number of random dimensions and model nonlinearities, with a very limited number of preliminary data. We note that, once the PCE's coefficients have been computed, 100,000 evaluations of the expansion would take, on the same Intel$^\circledR$ Core$^\text{TM}$ 2 Duo computer, about 80$\,$s, while the corresponding simulations with the dedicated real-time hardware would take about 46 days.

\section{Conclusions}\label{S:Conclusions}
We proposed a new method to compute polynomial chaos expansions, by means of a suitably defined convex optimization problem. The method can easily handle thousands of terms in the PCE, corresponding for example to stochastic dimensions of 15-20 with orders of 3-4. Bounds on the first and second order moments and on the values of the resulting PCE can also be explicitly included. We applied the approach to three examples in a broad range of different fields: in all cases, the derived PCEs, computed via a very low number of preliminary simulations,  accurately captured the process' statistics, despite the presence of nonlinearities and high stochastic dimensions. This aspect indicates that a quite small number of sampled simulations already contains sufficient information on the process, to derive an accurate PCE approximation. This method can be straightforwardly used in a large variety of applications, since it does not require any modification to the existing model, but just a small number of simulation runs.\\
$\,$\\

\textbf{Acknowledgments}\\
This research has received funding from the European Union Seventh Framework
Programme (FP7/2007-2013) under grant agreement n. PIOF-GA-2009-252284 -
Marie Curie project ``ICIEMSET'', from the Air Force Office of Scientific Research under the MURI award n. FA9550-10-1-0143, from the  National Science Foundation through Grants  ECCS-0835847 and ECCS-0802008 and from the Institute for Collaborative Biotechnologies through Grant DAAD19-03-D-0004 from the US Army Research Office.

\bibliography{Bibliografia}

\end{document}